%% file: main.tex
\documentclass[final,5p,twocolumn]{elsarticle}

\usepackage{caption}
\usepackage{mwe}
\usepackage[dvipsnames,table,xcdraw]{xcolor}
\usepackage{lineno}
\usepackage[hidelinks]{hyperref}
\modulolinenumbers[5]
\usepackage{todonotes}
\usepackage{multirow}
\usepackage{appendix}
\usepackage[utf8]{inputenc}
\usepackage{graphicx}
\usepackage[caption=false]{subfig}
\usepackage[export]{adjustbox}
\usepackage{booktabs}
\usepackage{newfloat}
\usepackage[skip=0pt,belowskip=0pt]{caption}
\usepackage{balance}
\usepackage{comment}
\usepackage{amsmath,amssymb,amsfonts}
\DeclareCaptionType{Query}
\usepackage{enumitem}
\usepackage{pgf}
\usepackage{multirow}
\usepackage{rotating}

\journal{Journal of Systems and Software}

\usepackage{mweights}
\usepackage{stix}
\usepackage{ifthen}
\newboolean{showcomments}
\setboolean{showcomments}{true}

\ifthenelse{\boolean{showcomments}}
  {\newcommand{\nb}[2]{
    \fbox{\bfseries\sffamily\scriptsize#1}
    {\small$\blacktriangleright$\textit{#2}$\blacktriangleleft$}
   }
  }
  {\newcommand{\nb}[2]{}
  }

\usepackage{xspace}

\newcommand\topic{architectural support for SP within CSE\xspace}

\newcommand\searchDate{\textcolor{BrickRed}{March 1st, 2022}\xspace}
\newcommand\initialPapers{\textcolor{BrickRed}{215}\xspace}
\newcommand\initialRefinedlPapers{\textcolor{BrickRed}{195}\xspace}
\newcommand\titleAbsPapers{\textcolor{BrickRed}{74}\xspace}
\newcommand\conflictPapers{\textcolor{BrickRed}{32}\xspace}
\newcommand\snowballingPapers{\textcolor{BrickRed}{86}\xspace}
\newcommand\snowballingRefinedPapers{\textcolor{BrickRed}{11}\xspace}
\newcommand\snowballingDate{\textcolor{BrickRed}{March 2022}\xspace}

\newcommand\fullPapers{\textcolor{BrickRed}{85}\xspace}
\newcommand\SelectedFullPapers{\textcolor{BrickRed}{66}\xspace}

\usepackage{mdframed}
\usepackage{ntheorem}
\theoremstyle{nonumberplain}

\newmdtheoremenv[%
  linecolor=violet,
  linewidth=2pt,
  rightline=false,
  leftline=false]{figrev}{}
  
\theorembodyfont{\upshape}
\newmdtheoremenv[%
  linecolor=violet,
  linewidth=2pt,
  rightline=false,
  leftline=false]{tabrev}{}

\newcommand{\rev}[1]{{\leavevmode\color{blue}{#1}}}
\newcommand{\revnew}[1]{{\leavevmode\color{black}{#1}}}

\newcommand{\SoBigDataITHack}{European Union - NextGenerationEU - National Recovery and Resilience Plan (Piano Nazionale di Ripresa e Resilienza, PNRR) - Project: “SoBigData.it - Strengthening the Italian RI for Social Mining and Big Data Analytics” - Prot. IR0000013 - Avviso n. 3264 del 28/12/2021\xspace}

\begin{document}
\begin{frontmatter}

\title{Architectural Support for Software Performance \\in Continuous Software Engineering: A Systematic \\Mapping Study}



\author[unite]{Romina Eramo\corref{mycorrespondingauthor}}
\ead{reramo@unite.it}
\cortext[mycorrespondingauthor]{Corresponding author}

\author[charles]{Michele Tucci}
\ead{tucci@d3s.mff.cuni.cz}

\author[univaq]{Daniele Di Pompeo}
\ead{daniele.dipompeo@univaq.it}

\author[univaq]{Vittorio Cortellessa}
\ead{vittorio.cortellessa@univaq.it}

\author[univaq]{Antinisca Di Marco}
\ead{antinisca.dimarco@univaq.it}

\author[oulu,tampere]{Davide Taibi}
\ead{davide.taibi@oulu.fi}

\address[unite]{University of Teramo, Teramo, Italy}
\address[univaq]{University of L'Aquila, L'Aquila, Italy}
\address[charles]{Charles University, Prague, Czech Republic}
\address[oulu]{University of Oulu, Oulu, Oulu, Finland}
\address[tampere]{Tampere University, Tampere, Finland}

\begin{abstract}
The continuous software engineering paradigm is gaining popularity in modern development practices, where the interleaving of design and runtime activities is induced by the continuous evolution of software systems. In this context, performance assessment is not easy, but recent studies have shown that architectural models evolving with the software can support this goal.
In this paper, we present a mapping study aimed at classifying existing scientific contributions that deal with the architectural support for performance-targeted continuous software engineering. 
We have applied the systematic mapping methodology to an initial set of \initialPapers potentially relevant papers and selected \SelectedFullPapers primary studies that we have analyzed to characterize and classify the current state of research.
This classification helps to focus on the main aspects that are being considered in this domain and, mostly, on the emerging findings and implications for future research.
\end{abstract}


\end{frontmatter}


\input{intro}

\input{related}

\input{methodology}

\input{results}
\input{threats}
\input{conclusion}


\section*{Acknowledgments}
\noindent This work was partially supported by the Adoms grant from the Ulla Tuominen Foundation (Finland) and the MuFAno grant from the Academy of Finland (grant n. 349488), the AIDOaRt project grant from the ECSEL Joint Undertaking (JU) (grant n. 101007350),  Territori Aperti project funded by Fondo Territori, Lavoro e Conoscenza CGIL CISL UIL, and SoBigData RI project funded by H2020-INFRAIA-2019-1 EU (grant n. 871042).
Daniele Di Pompeo is supported by the \SoBigDataITHack. Michele Tucci is supported by the OP RDE project No. CZ.02.2.69/\-0.0/\-0.0/\-18\_053/\-0016976 ``International mobility of research, technical and administrative staff at Charles University''.

\bibliography{biblio}

\appendix
\section{The Selected Papers}
\input{primary_studies}
\section{\rev{Data Extraction}}
\noindent
Table \ref{tab:data-extraction-results} presents an overview of the results obtained from the data extraction. In particular, it shows the list of the selected papers (SP) and their evaluation with respect to the selected keywords (note that, black squares represent fully investigated keywords, while white squares represent partially investigated keywords). In the next sections, we report the results of paper categorization for sake of answering the research questions of our study. For each research question, we discuss the results obtained outlining some implications for researchers and practitioners working in software architectures, performance engineering, and continuous software engineering. 

\input{table_data_extraction}
\section{Keywords definition}\label{sec:appendix}
In this appendix we defines the keywords we used during the data extraction. 
\input{table_keywords.tex}

\end{document}

%% file: intro.tex
\section{Introduction}\label{sec:intro}
\noindent
Continuous software engineering (CSE) is a promising software process that interleaves business strategy (i.e., requirement engineering), development, and operations on a continuum. It aims to produce a better software product and create more successful implementations that satisfy the relevant requirements and constraints. Similarly, the recent emphasis on DevOps recognizes that the integration between software development and its operational distribution must be continuous.
DevOps improves end-to-end collaboration between the stakeholders, development, and operations teams. In addition, they have been successfully employed in disciplines such as security and testing. 
Software performance (SP) is an essential quality aspect for the adoption and success of a software system. Researchers and industry practitioners have identified the importance of integrating performance engineering practices in continuous development processes in a timely and efficient way \cite{3053600.3053636}. However, current software performance engineering methods are not tailored for environments using CSE processes and practices are lagging \cite{arxiv.1808.06915, Kudrjavets2022}.

Although SP is a non-functional property related to the platform on which the software is deployed, performance assessment, in the last two decades, has been mainly estimated at the design level through methods, such as software architecture (SA) \cite{BrosigHK11,book,MartensKBR10}. 
SAs can be transformed into performance models, whose indices can be exploited to compare SA alternatives. Such a design-time performance assessment does not extensively consider several aspects of the target platform characteristics. 
However, these early-stage comparative analyses that show differences evident in the alternative results certainly support architects to make decisions with an enhanced view of their performance effects. 

The rise of the continuous engineering paradigm has substantially changed in the last decade of the software process. 
More often, it is nowadays required that software engineering follows a continuous loop between the running code and design models such that these two sides of the process can reciprocally feed each other \cite{CSE}. 
For example, runtime data can be collected from execution traces to feed software models. 
Software models are then aimed at checking either functional or non-functional properties. 
The analysis of software models in the context of incoming execution scenarios can suggest just-in-time refactoring/adaptation actions that keep the software behavior acceptable when these scenarios occur \cite{DBLP:conf/icsa/MazkatliMGK20,DBLP:conf/wosp/SpinnerWK16,DBLP:journals/jss/SpinnnerGEK19}. 
In the context of CSE processes, architectural models appear to have gained relevance, among others, for supporting performance-related decisions \cite{DBLP:conf/icsa/ArcelliCPET19}. 

Despite rising interest in embracing the continuous architecting approaches and performance engineering practices, there has been a little consensus in the literature on the appropriateness of different performance engineering techniques that can be used in a continuous engineering process. 
A limited number of studies that consider continuous engineering in some specific aspects of self-adaptive systems and microservices have been published ~\cite{Koziolek10, BeckerLB12, DBLP:conf/closer/PahlJ16, DBLP:journals/corr/abs-1908-10337, DBLP:conf/xpu/JabbariAPT16}.

However, current CSE and DevOps practices focus on rapid delivery, minimizing time to release for new features, mitigating risks, driving new efficiencies, and establishing a continuous delivery pipeline. Efficient and automated performance engineering tools are critical and pose relevant challenges in accomplishing this mission. Thus, there is still a need for a study that systematically investigates all the key publications on this topic and identifies possible performance engineering techniques applicable to continuous engineering processes. 

In this study, we conduct a mapping study of the existing literature~\cite{PETERSEN2015} to investigate the contributions of the scientific community to \topic.
Furthermore, the study aims to characterize and classify the current research scenario to better structure our understanding of the topic and identify research directions worth investigating in this domain soon. 


The main contributions of this study include:
\begin{itemize}
    \item A reusable framework for classifying, comparing, and evaluating solutions, methods, and techniques specific to architectural support for software performance in continuous software engineering;
    \item A systematic map of the state of research in the domain of architectural support for SP in CSE in terms of the performance areas, domains, addressed problems, and adopted instruments;
    \item A discussion of the emerging trends, gaps in the literature, and their implications for future research.
\end{itemize}

The remainder of this paper is organized as follows: In Section \ref{sec:related}, we provide a background review and compare the existing literature. In Section \ref{sec:method}, we define our target question and illustrate the process that we have adopted to conduct the mapping study; In Sections \ref{sec:results:pt}-\ref{sec:results:rq4}, we describe and analyze the results obtained to answer our target questions. In   Section~\ref{sec:threats}, we discuss the threats to validity of our study, and finally, Section \ref{sec:conclusion} presents the concluding remarks.

%% file: related.tex
\section{Background and Related Work}\label{sec:related}
\noindent
This section provides some background information and presents the synergies among the main concepts involved; other studies on related topics are also presented in this section.

\subsection{Main concepts}
\noindent
\textbf{Continuous Software Engineering (CSE)}. This refers to the capability to develop, release, and learn from software in rapid parallel cycles. This includes determining new functionality to build, evolve and refactor the architecture, develop the functionality, validate it, release it to customers and collect experimental feedback from the customers to inform the next cycle of development~\cite{TICHY2017173}. 

The definition of CSE is prone to interpretations and is often used in conjunction with other continuous activities that emerge during the entire software (engineering) lifecycle~\cite{CSE}. In particular, the activities considered in the \emph{development} phase are: continuous integration, continuous deployment/release, continuous delivery, and continuous verification/testing. Whereas, the \emph{operation} phase concerns the end of the process, where handover of the release is initiated; in this phase, particular attention is devoted to the continuous use of these systems, after the initial adoption, as well as continuous monitoring, to observe and detect compliance issues and risks. The most recent stand out of the \emph{DevOps}~\cite{DevOps} practices, which promote the integration between development and operations, confirms that these areas are closely interact to achieve CSE. Finally, a closer and continuous linkage between business management and software development functions is also necessary to benefit activities such as business planning; the \emph{BizDev}~\cite{CSE} phenomenon complements DevOps, integrating business management with software development and operations functions.

\noindent
\textbf{Software Performance (SP)}. This represents the entire collection of software engineering activities and related analyses used throughout the software development cycle, which are directed at meeting performance requirements~\cite{4221619}.
This field focuses on the quantitative evaluation of modern software systems (e.g., data-intensive, autonomous, distributed, adaptive, and embedded systems) and trade-offs between performance and other quality of service (QoS) attributes (e.g., security, reliability, and availability). In the last few decades, numerous performance engineering methods, methodologies, and techniques have been developed for system evaluation \cite{Merseguer17}.

SP assessment is a crucial task in software engineering to ensure that a new software release does not impair the user-perceived performance. Performance degradation can occur
in various forms, such as high response time latency, low throughput, and excessive resource utilization. 
Although these arguments would suggest that performance should be assessed on every change, recent studies on continuous engineering shows that it is not
standard practice yet \cite{3338982}.

\noindent
\textbf{Software Architecture (SA)}. 
The SA of a software system is the structure or structures of the system, which comprise software components, the externally visible properties of those components, and the relationships among them~\cite{SA2003}.
SA is often the first design artifact to represent decisions on how the requirements of all types are to be achieved. It shows the correspondence between the requirements and the constructed system, thereby providing a rationale for the design decisions~\cite{Hasselbring2018}. 
The design of the overall system structure, particularly in large and complex systems, is an essential factor. For instance, performance depends largely upon the complexity of the required communication, coordination, or cooperation among the different components, particularly in complex distributed systems.

The need for SA evaluation is based on the realization that software development, similar to all engineering disciplines, is a process of continuous modeling and refinement. Detecting architectural problems before the bulk of the development work is completed allows re-architecting activities to take place in due time without to rework what has already been done. At the same time, tuning activities enhance and maintain the SP during the software lifetime \cite{smr.337}.

\begin{figure}[htbp]
    \centering
    \includegraphics[width=1\linewidth]{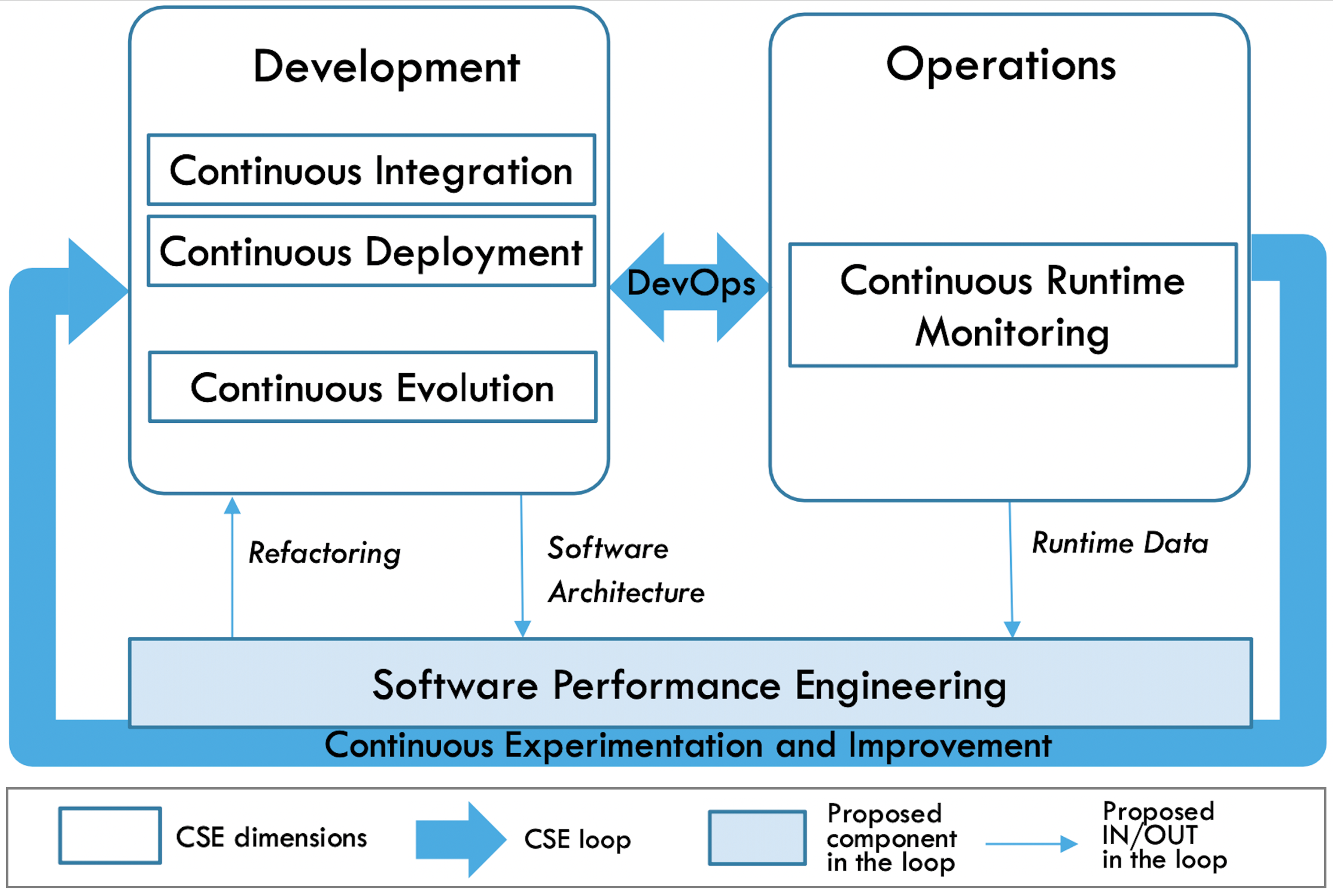}
    \caption{Overview of the considered context}
    \label{fig:CSE+SPE}
\end{figure}

\noindent
\textbf{Synergies between CSE, SP, and SA.}
\noindent
CSE involves several challenges in terms of SA evolution and the detection/resolution of problems related to software quality attributes, such as performance~\cite{CSE}.
Software development, in practice, concerns the continuous evolution of software, primarily owing to new incoming requirements. It is possible that SA is not adequate to embed new required functionalities, thus imposing a heavy and complex software evolution. This is compounded by situations in which the original developers are no longer available. 
In such cases, the system maintainability  is strongly related to its architecture \cite{978-3-030-52991-8_2}. 

With continuous engineering practices, developers have greater control and visibility of defects and to access the quality attributes states, enabling them to remedy any potential issues during the system development. 
Interactions between design-time and run-time in software engineering allow for dynamic adaptation and ensure  non-functional properties and end-user expectations \cite{BruneliereEGBBG18}. %
Notably, continuous monitoring is considered an enabler for the early detection of QoS problems, such as performance degradation \cite{HasselbringH20}.

Figure \ref{fig:CSE+SPE} illustrates the context of this study. We considered the holistic view of the CSE proposed in \cite{CSE} and tailored the figure by focusing only on the continuous activities that are central to SP (in the white rectangles) and by adding the specific task of performance assessment (in the light blue rectangle).

This figure also shows the bridge artifacts between CSE and SP. SA and runtime data are output artifacts of CSE that feed THE SP, whereas refactoring enters the CSE. SA is the abstraction that represents the best trade-off between model complexity and expressiveness and allows the assessment of the performance characteristics of a (software) system. Runtime data represents the running system that provides all parameters to set the performance model defined by the SA. Refactoring consists of suggestions on how to change the software system to solve or mitigate performance degradation.

Based on the above tailoring, we formulate a query to extract the literature object of our analysis.

\subsection{Related works}
\noindent
In this section, we discuss secondary studies that somehow address the role of SA and SP in the CSE paradigm.

Koziolek~\cite{Koziolek10} conducted a holistic literature review classifying the approaches concerning performance prediction and measurement for component-based software systems based on studies published from 1999 to 2009. These approaches introduce specialised modelling languages for the performance
of software components and aim to understand the performance of a designed architecture instead of code-centric performance fixes. The review acknowledges the limited support for the runtime life-cycle stage of software components and the lack of consensus on the performance modeling language.  In fact, none of the reviewed approaches was ready to gain widespread use due to limited tool support, fundamental in the case of CSE. The surveyed methods support modelling the runtime life-cycle stage of software
components only in a limited way (i.e., they only included the workload and the usage profile modeling of the component-based system at runtime). For continuous performance improvement, dynamic software refactoring is of paramount importance. However, the reviewed primary studies in ~\cite{Koziolek10} partially supported dynamic and automated mechanisms for CSE for performance aspects. Furthermore, online performance monitoring at runtime was not fully combined with modelling techniques to react on
changing usage profiles and deployment environments. 
As an extension of the review published by Koziolek~\cite{Koziolek10}, we present an updated analysis of the literature in our study including papers published until February 2022. Differently from ~\cite{Koziolek10}, our work focuses on publications investigating performance engineering methods that can be applied in the context of CSE and considering approaches applicable to all kind of systems without limiting our study to component-based software systems.

In a subsequent holistic literature review, Becker et al.~\cite{BeckerLB12} specifically investigated model-driven performance engineering approaches for self-adaptive systems based on studies published from 2004 to 2011. The authors provided a thorough classification scheme, presented as a feature diagram. They distinguished between the reactive and proactive adaptation strategies, and they derived two main categories of adaptation: design-time and run-time.

\revnew{Self-adaptation is the ability of the system to decide autonomously (i.e., without or with minimal human intervention) how to adapt to accommodate changes in its context and environment, 
and to manage the uncertainty in the environment in which the software is deployed, and during the execution \cite{weyns2020introduction}. Self-adaption is enabled since the self-adaptive systems use an explicit representation of their own structure, behavior, and goals \cite{LemosGGGALSWBBB13}. Recent efforts have been devoted to investigating motivation and the application of self-adaptation in practice~\cite{9799817}.
In this context, CSE defines a continuous engineering process needed to quickly respond to market and customer new requirements, i.e., to build solutions that much more accurately align with dynamic customer needs \cite{bosch2014continuous}.
A number of continuous activities (such as continuous monitoring, continuous integration, and so on) are part of an overall CSE~\cite{fitzgerald2017continuous}.
CSE and (self-)adaptation are two different run-time mechanisms in the sense that self-adaptation is the ability of a system to manage changes and uncertainty, while CSE is a dynamic process that continuously engineers the system allowing to add new features, functionalities, and abilities, or new smarter implementations of them.
}


While the aforementioned study (i.e.,~\cite{BeckerLB12}) consistently outlines performance engineering in self-adaptation targeting model-driven performance approaches, we seek to extend the area of interest to a more general interleaving of runtime knowledge and architectural models in CSE, without limiting the study to model-driven performance approaches. 

Recently, different studies have covered the different aspects of CSE~\cite{DBLP:conf/closer/PahlJ16,DBLP:journals/corr/abs-1908-10337,DBLP:conf/xpu/JabbariAPT16} in several contexts. However, differently from our work, they did not specifically consider SP engineering. Pahl et al.~\cite{DBLP:conf/closer/PahlJ16} have presented a systematic mapping study of 21 papers published from 2014 to 2015 to identify, taxonomically classify, and systematically compare the existing research body on microservices and their application in the cloud, by positioning them within the context of continuous development. 
Taibi et al.~\cite{DBLP:journals/corr/abs-1908-10337} have presented a systematic mapping study of continuous architecture with microservices and DevOps, and included 23 studies published from 2014 to 2017 in their investigation. They provided an overview of the architectural styles of microservices applications, highlighting the advantages and disadvantages of each style. However, no consideration was given to non-functional properties, such as performance. Jabbari et al.~\cite{DBLP:conf/xpu/JabbariAPT16} presented a systematic mapping study on the classification of DevOps and included 49 papers published from 2011 to 2016. They investigated how DevOps was exploited during software development processes. They found that few primary studies exploited model-driven engineering techniques and focused on quality assurance.

Finally, Bezemer et al.~ \cite{Bezemer2019} conducted an industrial survey to gain insights into how performance is addressed in industrial DevOps settings. In particular, they have investigated the frequency of executing performance evaluations, the tools being used, the granularity of the performance data obtained, and the use of model-based techniques. 

In contrast to the aforementioned papers, in this paper, we execute a systematic mapping study that investigates how performance is assessed in the context of CSE by providing a classification schema able to classify primary studies concerning research areas, addressed target problems, provided contributions, devised methodologies, studied performance indices, and type of used data. Unlike other related work, the focus of this study is the combination of SA and SP within CSE.

%% file: methodology.tex
\section{Research Method}\label{sec:method}
\noindent
To gain insights into the current research practices on the \topic, we conducted a systematic mapping study of the literature based on the guidelines proposed by Petersen et al.~\cite{PETERSEN2015}, and the ``snowballing'' process defined by Wohlin~\cite{Wohlin2014}.

\begin{figure*}[htbp]
    \centering
    \includegraphics[width=.7\linewidth]{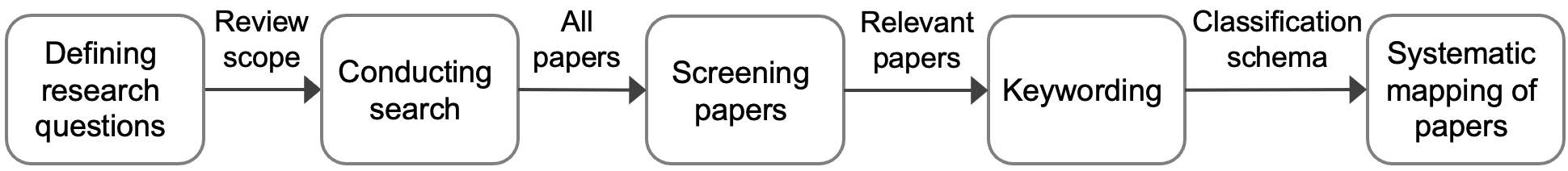}
    \caption{Mapping process}
    \label{fig:process}
\end{figure*}

The process adopted in this study consists of five steps, as shown in Figure~\ref{fig:process}. In the first step, we \emph{define the research questions} and identify the scope of the review to be incorporated in the next steps. Subsequently, we \emph{conduct a literature search} to retrieve a list of relevant publications that are then selected by applying the inclusion and exclusion criteria in the \emph{papers selection} step. The selected publications are the input for the \emph{data extraction}, where we categorize the relevant publications by considering their full text. As output, we obtain a classification schema that is used as input for \emph{mapping the data retrieved from the papers} to the questions. 

In the following section, we describe the five aforementioned steps of the mapping process. In Sections~\ref{sec:results:pt}-\ref{sec:results:rq4}, we present the results of the analysis and mapping of the papers.
Moreover, to simplify the replicability of this study, a complete replication package is made publicly available~\cite{ReplicationPackage}. 

\subsection{Defining research questions}\label{sec:process-rqs}
\noindent
To investigate the contributions of the scientific community on the \topic, we formulated the following research questions: 

\paragraph{RQ1: What \textbf{research areas} and \textbf{target systems} have been investigated?} 

The aim of this research question is two-fold: \emph{i)} to highlight the research areas that are focused on providing solutions in this field; and \emph{ii)} to extract the subject systems on which the application or technique is intended to apply (we refer to this by the term ``target system" \cite{Tryptych, domainSE}).

The rationale for this RQ is strongly related to the goal of this study and it aims to define how and what degree of performance engineering is exploited in CSE. It also determines which application domains have been considered in the selected studies. This helps us to understand the maturity of continuous performance assessment and to determine the applications for which performance is considered a key constraint.     

\paragraph{RQ2: What and how \textbf{performance problems} have been addressed?}

This research question focuses on the identification of the SP engineering problems targeted in a CSE process and the solutions proposed to address them. 

Several issues can be addressed in SP engineering, including requirement specification, modeling, analysis, prediction, and  suggestions to improve the software system performance. 
The rationale behind this RQ is strictly related to the identification of the SP target problems considered by the researchers and the related contributions proposed in the context of CSE.

\paragraph{RQ3: What \textbf{instruments} have been adopted?}
Several instruments can be used to address the performance issues. We partitioned these into three categories of keywords: input data, methodologies/techniques, and performance output measures/indices. The first category includes the types of data that are used to conduct the performance analysis, and spans from runtime data through requirements to software/performance models.  Examples of the the second category are patterns/anti-patterns recognition, performance prediction or testing. The third category aims to identify the target metrics, such as response time, throughput and network bandwidth.  This research question aims to identify which, and with what degree instruments have been applied in the context of CSE.

The rationale behind this question is strictly related to determining the characteristics and limits of the proposed solutions in the SP and CSE domains. 

\paragraph{RQ4: What are the gaps in current research \textbf{gaps} and the implications for future research?}

This research question combines the different viewpoints highlighted in the previous three RQs, and aims to identify contexts that have been hitherto most or least investigated. For example, how intensively has \emph{performance assessment} (as a problem) been investigated in the context of \emph{continuous monitoring} (as a research area) on \emph{distributed systems} (as a target)? 

\medskip

We are particularly interested in highlighting combinations that exhibit low intensities. We expect that some of these combinations to raise negligible interest and others to represent research gaps. Moreover, we are focused on identifying areas worth investigating in the near future.  



\subsection{Conducting search}\label{sec:process-search}
\noindent
The search process involves the identification of search strings, the outline of the most relevant bibliographic sources and search terms, the definition of the inclusion and exclusion criteria, and the query execution. 

\textbf{Search strings}. We defined the search keywords based on the PICO\footnote{PICO elements include: problem/patient/population, intervention/indicator, comparison, outcome~\cite{huang2006evaluation}.} terms~\cite{Kitchenham2007} in our questions structure.
As suggested by Kitchenham~\cite{Kitchenham2007}, the comparison and outcome terms cannot always be considered in software engineering if the research focuses on general investigation. hence, we extracted the keywords from the population and intervention terms.


\begin{table*}[!ht]
\footnotesize
\centering
\begin{tabular}{lp{12cm}}
\toprule
\textbf{Population} & \textbf{P Terms} \\
\midrule
Software Performance &  software performance\\
\bottomrule
\bottomrule
\textbf{Intervention} & \textbf{I Terms} \\
\midrule
Software Architecture &  software architecture \\
\midrule
Continuous Software Engineering &  DevOps, continuous integration, continuous deployment, continuous development, continuous improvement, DevOps, continuous evolution, continuous monitoring \\
\bottomrule
\end{tabular}
\caption{Definition of keywords}
\label{tab:SearchString}
\end{table*}

We refined the search terms and the related search strings to ensure that relevant studies were returned by combining the keywords and reviewing the titles and abstracts of the search results. The final set of keywords is listed in Table~\ref{tab:SearchString}. 
The resulting query was then adapted to the syntax of each bibliographic source. All the queries applied to the different bibliographic sources are reported in the replication package~\cite{ReplicationPackage}. 

\vspace{-0.5cm}
\begin{center}
\fbox{%
    \parbox{0.98\linewidth}{%
        \emph{(“continuous software engineering” OR “continuous integration” OR “continuous deployment” OR “continuous development” OR “continuous improvement” OR DevOps OR “continuous evolution” OR “continuous monitoring”) AND “software architecture” AND “software performance”}}}\par%
  \captionof{Query}{\label{query:baseline}Baseline search string.}
\end{center}

\textbf{Bibliographic sources}. We selected a list of relevant bibliographic sources following the suggestions of Kitchenham and Charters~\cite{Kitchenham2007}, as these sources were recognized as the most representative in the software engineering domain and were used in many reviews. The list includes: \textit{ACM Digital Library, IEEEXplore Digital Library, Scopus, and Springer Link}.

\textbf{Inclusion and exclusion criteria}. We defined the inclusion and exclusion criteria to be applied to the title and abstract (T/A) or to the full text (All), as reported in Table~\ref{tab:Criteria}.

\begin{table*}[!h]
\centering
\footnotesize
\begin{tabular}{lp{15cm}p{0.4cm}}
\toprule
\textbf{Criteria} &\textbf{Assessment Criteria} & \textbf{Step} \\ 
\midrule
\multirow{5}{*}{Inclusion} 
& The paper covers software performance engineering issues & All \\ \cline{2-3}
& The paper proposes model-based or architectural approaches for CSE/DevOps  or 
contributes to (self-) adaptation/refactoring targeted to software performance
& All\\ 
 \midrule
\multirow{6}{*}{Exclusion} 
& The paper is not fully written in English  & T/A \\ \cline{2-3}
& The paper is not peer-reviewed (i.e., blog, forum, etc.) & T/A\\ \cline{2-3}
& The paper is a duplicate (only consider the most recent version)& T/A \\ \cline{2-3}
& The paper is a position papers, book chapter or work plan (i.e., the paper does not report results) & T/A \\ \cline{2-3}
& The paper does not fully or partly focus on software performance & All\\ \cline{2-3}
& The paper does not fully or partly focus on software architecture or software engineering  & All \\ 
\bottomrule
\end{tabular}
\caption{Inclusion and exclusion criteria}
\label{tab:Criteria} 
\end{table*}

\begin{figure}[htbp]
    \centering
    \includegraphics[width=0.95\linewidth]{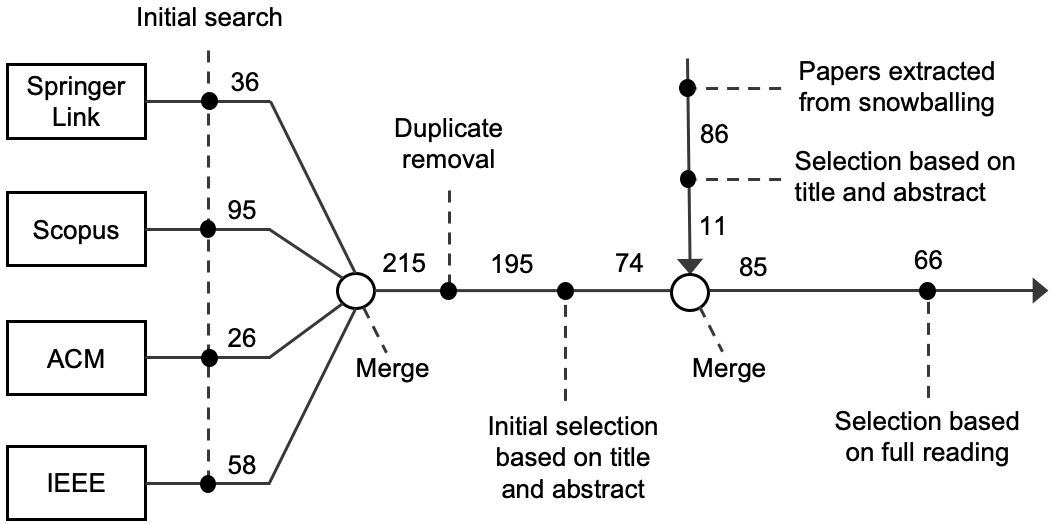}
    \caption{\label{fig:selection-process}Overview and numbers of search and selection process}
\end{figure}

\textbf{Search}. Finally, the search was conducted on \searchDate, and all the publications hitherto available were included. The application of the searching terms returned \initialPapers papers, which was the result of merging the papers from the bibliographic sources considered, as depicted on the left side of Figure~\ref{fig:selection-process}. Upon removing the duplicate papers, we obtained \initialRefinedlPapers papers.

We validated the search string with a ``golden set'' of papers that ensured that we did not leave out relevant works. 
The papers considered in the golden list were: 
\ref{bib:Garlan2004},
\ref{bib:Brosig2011},
\ref{bib:Ehlers2011a},
\ref{bib:Calinescu2011},
\ref{bib:Brunnert2017}, and
\ref{bib:Bezemer2019}.

\subsection{Papers selection}\label{sec:process-selection}
\noindent
After obtaining the initial set of papers, we applied the selection process described in this section. An overview and numbers of this process is depicted in Figure~\ref{fig:selection-process}. 


\textbf{Testing the applicability of the inclusion and exclusion criteria.} Before applying the inclusion and exclusion criteria, all the authors tested their applicability iteratively, on a subset of 20 randomly selected papers.
Based on the disagreements, and on a shared discussions, we clarified the inclusion and exclusion criteria.

\textbf{Applying the inclusion and exclusion criteria to the title and abstract.} The refined criteria were applied to the remaining \initialRefinedlPapers papers (Table~\ref{tab:Criteria}). We have included papers that meet all the inclusion criteria and excluded those that meet any of the exclusion criteria. 
Each paper was read by two authors; in the case of disagreement, a third author helped to resolve the disagreements.
For \conflictPapers papers, the authors discussed and cleared possible disagreements. Out of the \initialRefinedlPapers initial papers, we included \titleAbsPapers papers based on the title and abstract.
The inter-rater agreement before the third author was involved was 0.75, obtained using Cohen's kappa coefficient, which indicated a substantial agreement between the authors~\cite{interrater}.

\textbf{Snowballing.} We performed the snowballing process~\cite{Wohlin2014}, by considering all the references presented in the retrieved papers and evaluating all the papers referencing the retrieved papers, which resulted in one additional relevant paper. We applied the same process to papers retrieved from the initial search. A
snowballing search was conducted in \snowballingDate. We identified \snowballingPapers potential papers, but only \snowballingRefinedPapers were included (after applying the inclusion and exclusion criteria to the title and abstract) in order to compose the final set of publications that were subjected to full reading and data extraction.

\textbf{Full reading.} 
The screening of the remaining \fullPapers papers was performed independently by two authors. We ensured that the papers were randomly assigned such that each author had a similar number of papers assigned. Moreover, we permuted the assignments to enable a good balance between each pair. 

We read the \fullPapers papers in full and applied the criteria defined in Table~\ref{tab:Criteria}. 
To improve the reliability of our study~\cite{Wohlin2013}, we sought the services of a third author in two 
papers to reach a final decision.
In this case, the inter-rater agreement before the third author was involved was strong (Cohen's kappa coefficient = 0.94; almost perfect agreement).
Based on this process, we selected a total of \SelectedFullPapers papers for the review.

 \subsection{Data Extraction and Analysis }
\label{sec:process-extraction} 
\noindent
%
To ensure a rigorous data extraction process and to ease the management of the extracted data, a well-structured classification framework was rigorously designed, as explained in this section. 

To answer our RQs, we extracted a set of information from the \SelectedFullPapers selected papers. Notably, we defined the main concepts and corresponding data in our study by following a systematic process called \emph{keywording}. The goal of this process is to effectively develop a classification scheme so that it fits the selected papers and considers their research focus into account \cite{Petersen2008}. In particular, we
identified the codes for our coding schema using a semi-automated process in the following two steps:
\begin{enumerate}
    \item \textit{Automatic identification of the most recurrent keywords in the papers}.
    We used natural language processing (NLP) techniques to automatically identify the keywords that were most frequently mentioned in the abstracts of the selected papers.
    We started by collecting the abstracts from a single dataset that constituted the text corpus for the processing.
    The corpus was pre-processed in two phases: \emph{noise removal} and \emph{normalization}.
    In the \emph{noise removal} phase, we performed an initial clean-up by converting the text to lowercase and by removing punctuations, tags, special characters, and digits.
    We then applied two normalization techniques: \emph{stemming} to remove suffixes and \emph{lemmatization} to group together words having the same root.
    As a final pre-processing step, we removed the prepositions, pronouns, and conjunctions.
    Thus, we created a vector of words counts by deriving a \emph{bag-of-words} model for the text.
    In this model, words order and grammar information are not considered because the entire text is represented by the multiset of its words from which one can derive their multiplicity.
    The vector of words counts was then used to obtain the 50 most frequent single words and two words (bi-grams) and three words (tri-gram) combinations.
    
    \item \textit{Manual refinement of the keywords.} We refined our collection of keywords and concepts by reading the  abstract of each paper. We combined together keywords from different papers to develop a high level understanding of the nature and contribution of the research. This helped us to define a set of categories of keywords that is representative of the research questions. However, the paper abstracts were too limited to define all meaningful keywords. Therefore, we thoroughly examined all the sections of the papers to consolidate our classification schema. We performed a double round of reviews by shuffling reviewers (among the authors of the paper) after the first round. Finally, upon obtaining a consolidated set of keywords, we have re-organized the original categories to obtain the final classification used hereafter.
\end{enumerate}

We assigned each author a set of 10 randomly selected papers, to validate the coding schema and keywords, and to ensure a common understanding among the researchers. Subsequently, we discussed on the results of the coding and possible inconsistencies, and we finalized the schema. 

The resulting classification framework is presented in 
Table~\ref{tab:DataExtraction2}. 
It comprises seven categories, with groupings of pertinent extracted keywords. A detailed description of each keyword is provided in \ref{sec:appendix}.
Each category addresses the corresponding research questions by using the metrics described in details below.  

\input{table_2}

For \emph{RQ$_1$}, we extracted the information on the research area and target system. 
Both research areas and targets may be either fully or partially investigated. The primary goals of a paper are included in the fully-investigated research areas and targets.

In a partially-investigated research area, either the primary goals of the paper are considered, or it is a secondary area that supports the primary goals (while a partially-investigated target implies that the study can support that targeted system, even if it is not described as central in the paper). 
As an example, we considered papers, such as \ref{bib:Bernardi2018}, where \emph{continuous software engineering} has been partially investigated because the paper mainly focuses on \emph{software performance engineering}. 
It is important to note that papers might fully investigate one area and partially investigate another one, and therefore there might be more than one research area assigned to each paper. 

Regarding the problems addressed in the selected papers for \emph{RQ$_2$}, we extracted information on the primary and secondary problems and contributions reported by the selected  papers. As an example, we considered \emph{performance requirement} as the primary problem and \emph{quality of service} as the secondary problem for paper~\ref{bib:Yasaweerasinghelage2019}, while we considered \emph{performance analysis approach} as main contribution and \emph{performance modelling approach} and \emph{performance prediction approach} as the secondary contributions for the paper \ref{bib:Trubiani2018}. 
The same approach was applied also for \emph{RQ$_3$}.

After extracting all the information from the selected papers, we analyzed the data by counting the number of papers obtained for each data group and metrics (see Table~\ref{tab:DataExtraction2}). 
Therefore, for (\emph{RQ$_1$}), we counted the number of papers that fully investigated a topic, partially investigated it or considered the topic (either fully or partially investigated it). 
Similarly, for (\emph{RQ$_2$}), we counted the papers that considered each research problem (primary, secondary, or both). 
Following the previous approach, (\emph{RQ$_3$}) was analyzed by counting the number of fully- and partially-evaluated indices, used data, and methods. 

For the first three RQs, we considered the individual keyword results. \emph{RQ$_4$} has been introduced to observe the results across diﬀerent keywords, with the goal of identifying contexts that have scarcely been investigated. To achieve this goal, we introduced bubble plots, which allowed for a straightforward comparison of how intensively a certain context was investigated in comparison to other contexts. In the plots, the size of the bubbles represents the number of papers that investigate a specific keyword at the intersection of a research area (x-axis) and domain (y-axis). The exact number of papers is annotated for each bubble. In this way, one can visually identify areas of the plot where no bubbles or only small bubbles are observed, thereby establishing combinations of research areas and target systems where a certain keyword appears to be seldom investigated in the considered literature. Moreover, we analyzed the combination of the previous results and future improvements and direction described in the selected papers to identify a set of implications for future research.

A complete list of the keywords and metrics used for the analysis is presented in Table~\ref{tab:DataExtraction2}.


%% file: table_2.tex
\begin{table*}[htbp]
\footnotesize
\begin{tabular}{lp{3cm}p{9cm}l}
\toprule
 \textbf{RQ}  &  \textbf{Categories}  &  \textbf{Keywords}  & \textbf{Metrics}                                                                                  \\ \midrule
  \multirow{6}{*}{RQ1} & Research area & \begin{tabular}[c]{@{}l@{}}Software performance engineering, Software architecture, \\ Continuous Software Engineering, DevOps, \\Continuous Monitoring, Agile software development.\end{tabular} & \begin{tabular}[c]{@{}l@{}}\multirow{5}{*}{\shortstack{\#Fully investigated topics (F) \\ \#Partially investigated topics (P) \\ \#Investigated topics (F, P)}}\end{tabular} \\ \cline{2-3} 
  & Target system   & \begin{tabular}[c]{@{}l@{}}{\color{blue}Embedded / CPS}, Cloud, Real-time, \\Distributed, Data intensive\\Software intensive, \\Component-based software / Microservices / SOA. \end{tabular}  & \\
\midrule
\multirow{6}{*}{RQ2} & Research problems &  \begin{tabular}[c]{@{}l@{}}{\color{blue}Performance evaluation/assessment}, \\Performance requirement, Quality of service, \\{\color{blue}Resource allocation / deployment}, Uncertainty.\end{tabular}    & \begin{tabular}[c]{@{}l@{}}\multirow{5}{*}{\shortstack{\#Main target \\ problem/contributions (M) \\ \#Secondary target \\ problem/contributions (S) \\ \#Main or Secondary target \\ problem/contributions (M, S)}} \end{tabular} \\ 
\cline{2-3} 
& Contributions     &  \begin{tabular}[c]{@{}l@{}}Performance analysis approach, Domain specific languages, \\Continuous engineering framework, \\ Performance modelling approach, Tool support, \\Performance prediction approach, Self-adaptation.\end{tabular} & \\ 
 \midrule
    \multirow{6}{*}{RQ3} & Methodologies     &   \begin{tabular}[c]{@{}l@{}}Performance model, Model based engineering /\\Model driven engineering (MBE/MDE), {\color{blue}Performance}\\ {\color{blue}antipattern / Root cause / Bottleneck detection}, \\ Performance prediction techniques, Performance analysis \\ techniques, Parametric dependency, \\{\color{blue}Performance testing / Load Testing / Benchmarking},\\ {\color{blue}Performance model generation/extraction}, Simulation, \\Machine Learning, (Multi-objective) Optimization.\end{tabular}  & \begin{tabular}[c]{@{}l@{}}\multirow{4}{*}{\shortstack{\#Fully evaluated \\ indices/data/methods (F) \\ \#Partially evaluated \\ indices/data/methods (P) \\ \#Fully or Partially evaluated \\ indices/data/methods (F, P)}}\end{tabular} \\ 
\cline{2-3} 
    & Indices (output)  &   \begin{tabular}[c]{@{}l@{}}Response time, Utilization, Throughput, Resource demand,\\ Network bandwidth, {\color{blue}Memory / Memory Leaks}.\end{tabular}    & \\ 
\cline{2-3} 
   & Used Data (input) &  \begin{tabular}[c]{@{}l@{}}{\color{blue}Runtime / Monitored}, Workload, Requirements, \\ Performance model, Software model, Data analytics.\end{tabular}    & \\ 
\midrule
\multirow{3}{*}{RQ4}  &  Research area and Target system combined with data from RQ2 and RQ3  &  All keywords of RQ1 combined with specific keywords of RQ2 and RQ3. & \begin{tabular}[c]{@{}l@{}}\multirow{3}{*}{\shortstack{\#Fully or Partially addressing \\a specific combination \\of keywords (F, P)}}\end{tabular}  \\ 
\bottomrule
\end{tabular}
\caption{Classification framework: Data Extraction, Keywords, and Metrics adopted for the analysis. Keywords in blue have been obtained in the manual keyword refinement step.} 
\label{tab:DataExtraction2} 
\end{table*}

%% file: results.tex



\input{results-intro}


\input{results-RQ1}\input{results-RQ2}\input{results-RQ3}\input{results-RQ4}\input{future}

%% file: results-intro.tex
\section{Results Overview}\label{sec:results:pt}



\begin{figure}
    \centering
    \includegraphics[width=.9\linewidth]{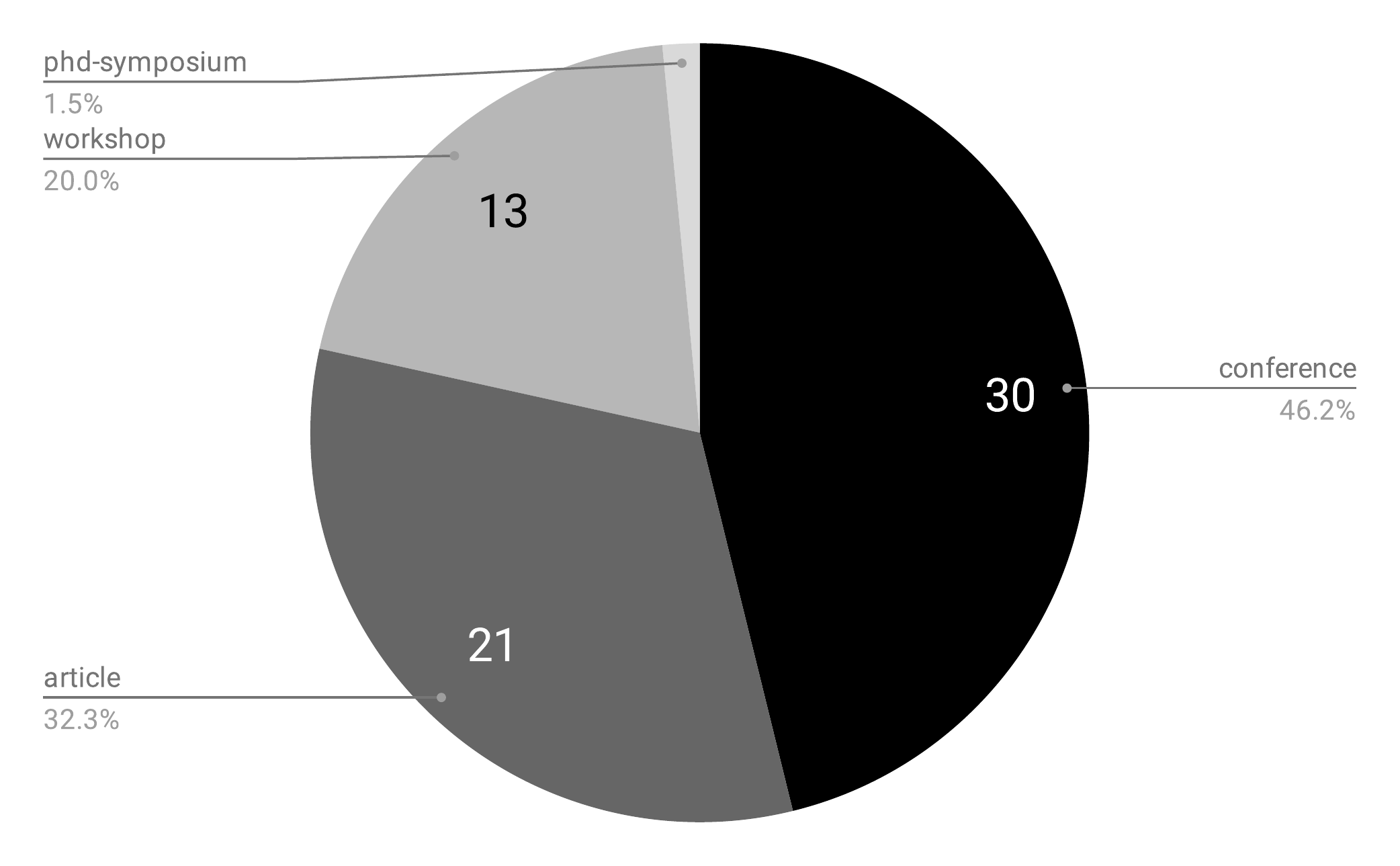}
    \caption{Selected paper types.}
    \label{fig:piechart-document-type}
\end{figure}

\begin{figure}
    \centering
    \includegraphics[width=.9\linewidth]{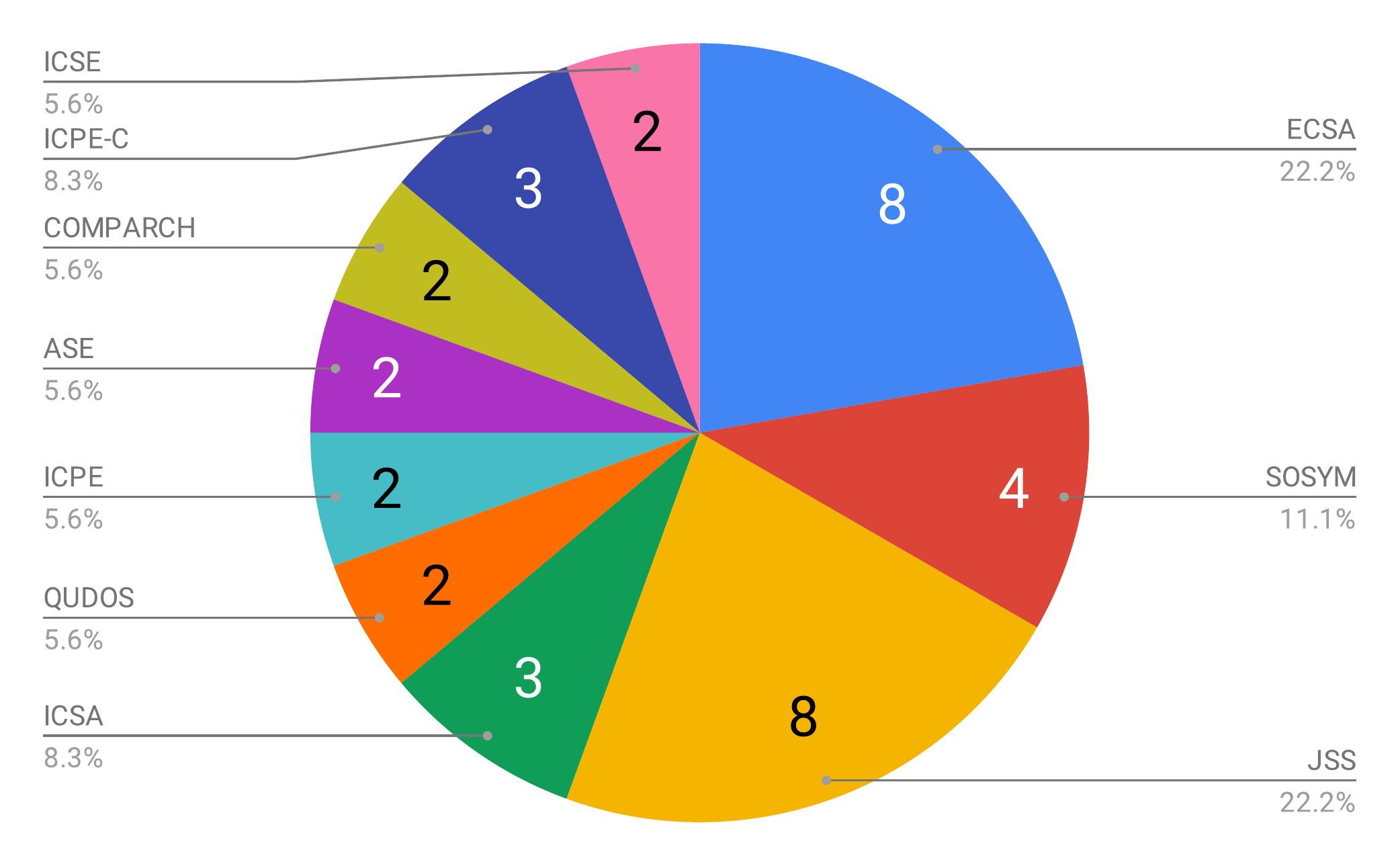}
    \caption{Selected paper per venues.}
    \label{fig:venues}
\end{figure}

\noindent
We selected \SelectedFullPapers peer-reviewed publications, including 21 (32.3\%) articles, 30 conference papers (46.2\%) and 15 workshop (or others satellite events) papers (21.5\%), as shown in Figure~\ref{fig:piechart-document-type}.
The selected papers were presented at 33 different venues. Figure~\ref{fig:venues} depicts the list of venues considered by at least two of the selected papers. 


The selected papers show a continuously growing interest on performance-targeted CSE between 2016 and 2022, while a very small number of publications have been published until 2015, which is in line with the fact that continuous development and DevOps have emerged only recently \cite{CSE}. 
Thus, we can gather that the intuition of supporting SP through continuous engineering solutions has been strengthened since 2015, although there has been a decrease in the number of publications in 2020 and 2021.

\begin{figure}[htpb]
    \centering
    \includegraphics[width=\linewidth]{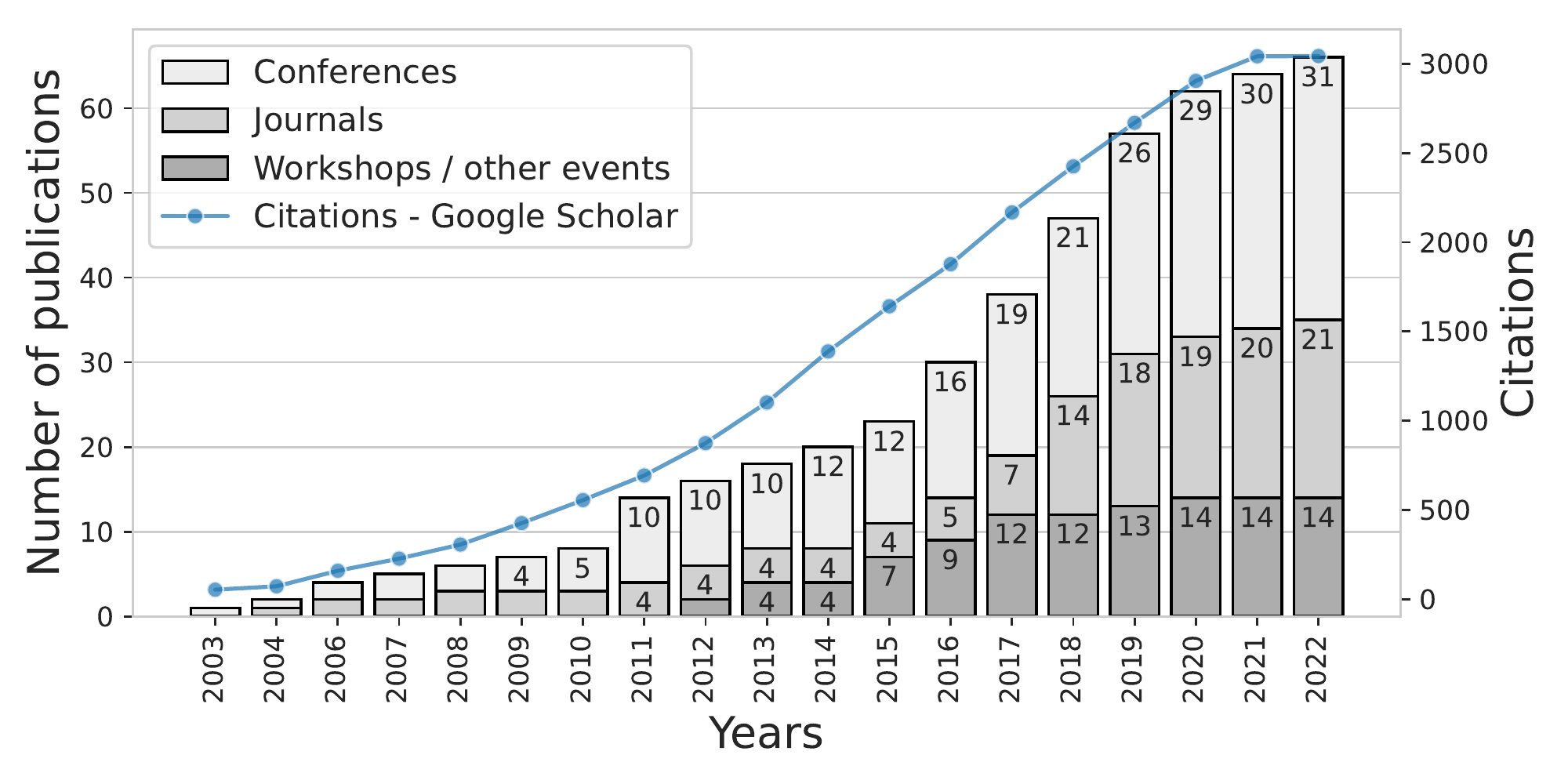}
    \caption{Cumulative number of primary studies and citations in each year, and by type of publication.}%
    \label{fig:pubs_by_year}
\end{figure}

The cumulative number of citations per year for the primary studies (Figure~\ref{fig:pubs_by_year}, blue line, source: Google Scholar)
highlights that the growing interest concerns not only the publications but also the number of citations obtained from the studies of the dataset. Moreover, citations have grown rapidly since 2016. It is worth noting that the entire dataset has a total of more than 3000 citations to date (i.e., they have more than doubled in the last six years). This result indicates an important growing interest in the context of this study.

Thus, although the DevOps and CSE domains are relatively young compared with performance engineering, significant contributions have been made in the last ten years and researchers are becoming increasingly active (Figure~\ref{fig:pubs_by_year}).



In the followin sectiong, we present the results of this study aimed at answering our research questions (see Section \ref{sec:process-rqs}). For each extracted piece of information, we report both the quantitative data and an interpretation of the results obtained.

%% file: results-RQ1.tex
\section{What research areas and target systems have been investigated (RQ$_1$)}
\label{sec:results:rq1}

\noindent
The  topic considered combines several aspects, as we discussed in Section~\ref{sec:related}, which may attract interest from different disciplines and for different scopes of research. To  provide an overview of this research topic, we describe the research areas focused on providing solutions and the target systems for which solutions have been developed.

\subsection{Research areas}
\begin{figure}[htbp]
    \centering
    \includegraphics[width=.95\linewidth]{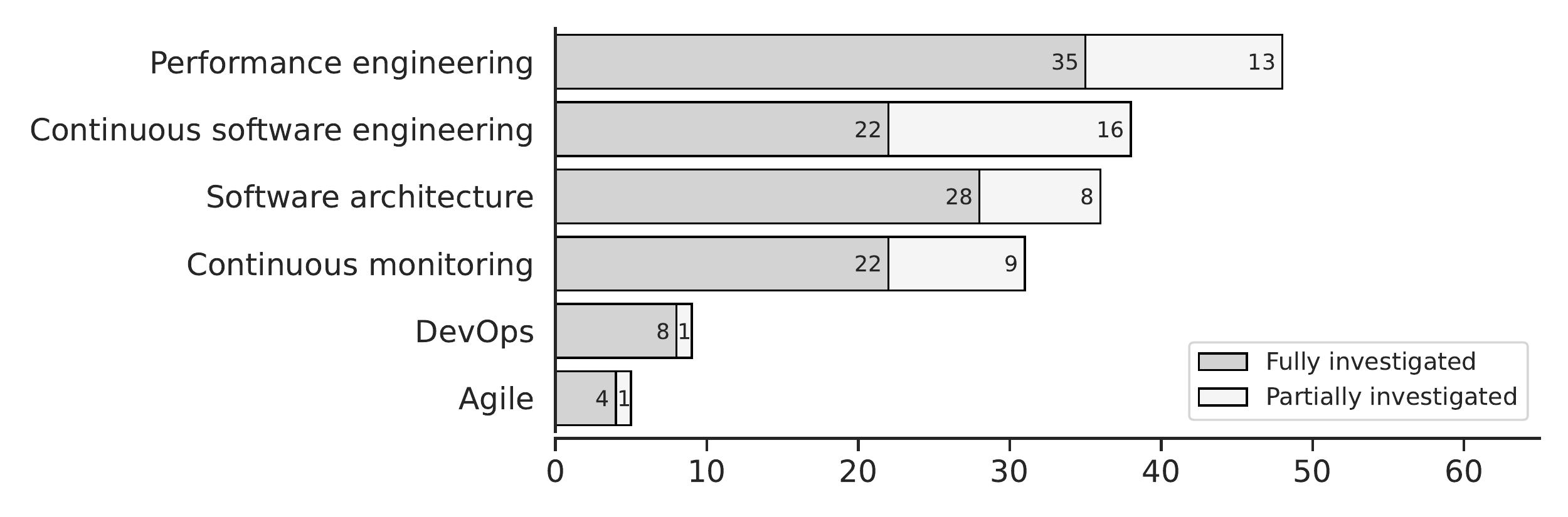}
    \caption{Research areas - results}
    \label{fig:research_area}
\end{figure}

\noindent
Figure~\ref{fig:research_area} depicts the principal \underline{\textbf{research area}}s that are focus areas of the selected papers  (each study may contribute to more than one area). The bar chart in the figure compares the identified research areas with respect to the number of papers. Although our topic is characterized by the dimensions showed in this figure, the fact that the selected papers contribute to these specific fields of research is not obvious. For instance, a paper may just use a specific performance evaluation technique without contributing in that area. For each research area, we stacked two diﬀerent bars to combine both the \emph{fully} and \emph{partially investigated} areas (as described in Section~\ref{sec:process-extraction}). 

As expected, the selected papers mainly focus on the areas of \emph{software performance engineering} ($48$ of \SelectedFullPapers papers) and \emph{software architecture} ($36$ papers), which is in agreement with the information on publication venues obtained in the publication trends.
On the $48$ papers, $28$ papers ($14$ fully) investigated both \emph{software performance engineering} and \emph{software architecture}, that is, they offered a specific contribution in architecture-based performance engineering of software systems. More specifically, of the $28$ papers, $14$ papers (fully or partially) intensively used architectural/software models; some of them were devoted to model-based software engineering (\ref{bib:Pooley2010}, \ref{bib:Voneva2020}, \ref{bib:Ehlers2011b},
\ref{bib:VonMassow2011}, \ref{bib:Yasaweerasinghelage2019}, \ref{bib:Brosig2011},  \ref{bib:Calinescu2011}, and \ref{bib:Spinner2016}), and others are focused on model-driven software engineering (\ref{bib:Arcelli2019}, \ref{bib:Vögele2018},  \ref{bib:Perez-Palacin2019}, \ref{bib:Heinrich2020}, and \ref{bib:Heinrich2016}).

Many studies have contributed to \emph{continuous software engineering} ($38$ papers). They considered at least one of the continuous dimensions introduced in Section \ref{sec:related}, that is,
continuous integration (8; \ref{bib:Wert2015}, \ref{bib:Pooley2010} \ref{bib:Falkner2018}, \ref{bib:Ehlers2011b}, \ref{bib:Giaimo2020}, \ref{bib:Willnecker2019}, \ref{bib:DeSanctis2020}, and \ref{bib:Mazkatli2020}),
continuous deployment (5; \ref{bib:Perez-Palacin2019}, \ref{bib:DeSanctis2020}, \ref{bib:Willnecker2016}, \ref{bib:Gerostathopoulos2016}, and \ref{bib:Heinrich2020}),
continuous development (3; \ref{bib:DeSanctis2020}, \ref{bib:Heinrich2020}, and \ref{bib:Stochel2012}),
continuous improvement (17; \ref{bib:Tsai2006}, \ref{bib:Pooley2010}, \ref{bib:Wert2015}, \ref{bib:Voneva2020}, \ref{bib:Bernardi2018}, \ref{bib:Falkner2018}, \ref{bib:Ehlers2011b}, \ref{bib:Giaimo2020},  \ref{bib:Willnecker2019}, \ref{bib:DelRosso2006}, \ref{bib:Perez-Palacin2019}, \ref{bib:Arcelli2019}, \ref{bib:Incerto2015}, \ref{bib:DeSanctis2020}, \ref{bib:Mazkatli2020}, \ref{bib:Incerto2017}, and \ref{bib:Trubiani2017}), and 
focused on CSE in general (10; \ref{bib:Arcelli2019}, \ref{bib:Incerto2015}, \ref{bib:Bezemer2019}, \ref{bib:DeSanctis2020}, \ref{bib:Mazkatli2020}, \ref{bib:Brebner2016},  \ref{bib:Keck2016}, \ref{bib:Walter2017}, \ref{bib:Heinrich2020}, and \ref{bib:Stochel2012}. 
Particular interest ($31$ papers)) was observed in regards to \emph{continuous monitoring} (hence, we decided to show it separately from the other continuous dimensions).

$Nineteen$ papers focused on the areas: \emph{software performance engineering}, \emph{software architecture} and \emph{continuous software engineering}, implying that they not only considered methodology and techniques of the areas but also provided a scientific contribution to these areas; thus, they were positioned in the context of this topic.  
Moreover, $15$ papers included a combination of \emph{software performance engineering}, \emph{software architecture} and \emph{continuous monitoring}. 

Only $nine$ of the selected papers covered \emph{DevOps} and only $five$ papers focused on the architectural support of performance engineering in the \emph{agile} development process. 
Of them, $three$ papers combined \emph{software performance engineering}, \emph{software architecture} and \emph{DevOps} (\ref{bib:Castellanos2019}, \ref{bib:Mazkatli2020}, and \ref{bib:Heinrich2020}), whereas $three$ papers combined \emph{software performance engineering}, \emph{software architecture} and \emph{agile} (\ref{bib:Pooley2010}, \ref{bib:Voneva2020}, and \ref{bib:Stochel2012}).

For each area, the relationship between \emph{fully-} and \emph{partially-investigated}
is proportional (i.e., a good percentage of the areas are fully investigated). However, only in regards to \emph{continuous software engineering}, the number of papers partially investigating the area increases. 

\subsection{Target systems}
\begin{figure}[htbp]
    \centering
    \includegraphics[width=.95\linewidth]{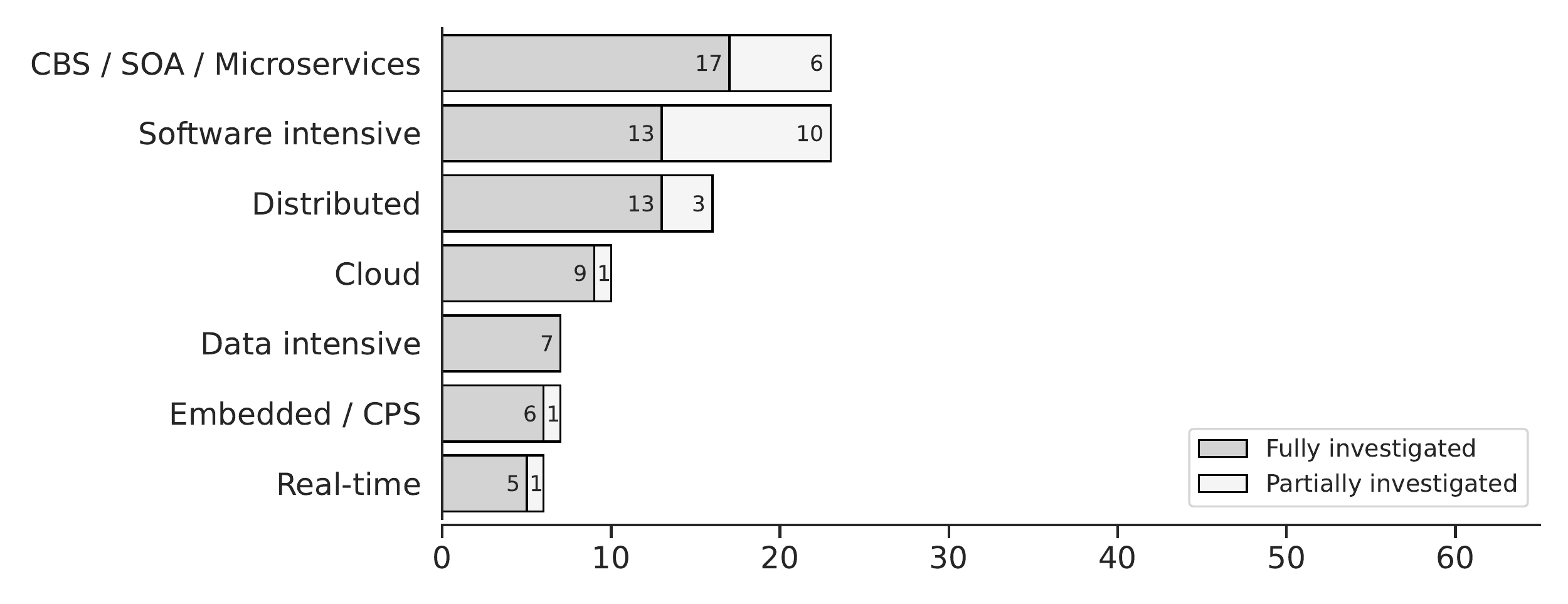}
    \caption{Target systems - results}
    \label{fig:domain}
\end{figure}

\noindent
\revnew{Figure~\ref{fig:domain} describes the specific type of systems targeted (as case studies) by selected papers. These keywords are not meant to be mutually exclusive, in that a system could be, for example, at the same time a distributed and real-time one. We have basically identified, in each paper, the main characteristics of systems to which the paper approach/solutions have been (or can be potentially) applied, where they have been unambiguously identified.}
The main focus is on \emph{software intensive systems} ($23$ papers), such as systems of systems (SoS) (\ref{bib:Falkner2018} and \ref{bib:Chiprianov2014}), automotive software (\ref{bib:Giaimo2020}), and 
business process management (\ref{bib:Pooley2010}). 

Similarly, a significant number of applications ($23$ papers) are in \emph{component-based systems (CBS)},   
and several papers focus on service-oriented architectures (SOA) (\ref{bib:Calinescu2011} and \ref{bib:ReinerJung2013}) and  microservices (\ref{bib:Cholomskis2018} and \ref{bib:Heinrich2020}).

In addition, \emph{distributed systems} ($16$ papers) have attracted considerable interest. However, a few of the paper cover \emph{embedded systems} (including \emph{cyber physical systems (CPS)} ($7$ papers) and \emph{real-time systems} ($6$ papers)). A new emerging application domain is related to big data and its management and analysis (\emph{data intensive systems}; $7$ papers).

Concerning the relationship between 
the \emph{fully-} and \emph{partially-investigated} targets,
it is observed that the larger targets (such as \emph{intensive software systems}) are represented by several papers, which are only partially placed in that context. This may depend on the fact that more than one target may be assigned to each paper. For instance, \ref{bib:Giaimo2020} fully covers the domain of \emph{CPS}, in particular automotive, and partially covers \emph{intensive software systems}). In other cases, some papers do not offer a solution dedicated to a specific target but are placed in a more general manner, as in case of \ref{bib:Incerto2017}.

\subsection{Discussion}
\label{sec:discussion:rq1}
\noindent
Significant attention has been paid to continuous monitoring of data to realize continuous development, delivery, and integration, which improve system performance at the level of SA. As the monitored data enable performance analysis, continuous monitoring represents a fundamental capability to provide CSE. In general, the results confirm that in the areas of SA and SP, CSE is an emerging topic that is progressively gaining the interest of researchers and practitioners. 
Existing conferences and systematic reviews on {DevOps} suggest that software engineering researchers have a strong interest in this topic. Despite this, only a few papers focuses on {DevOps},underling an interesting gap to be addressed by the research community. However, the limited number of articles on {agile} is an expected result. Even as a precursor to DevOps, agile development is more code-focused and produces less documentation (e.g., software/design models), disabling SA-based SPE.

Although a large number of selected papers have been fully identified as contributors to the CSE field, the number of papers that partially investigate this area has increased. Beyond that, several selected papers have not been placed within the context of CSE or DevOps, which can be attributed to the fact that these papers do not explicitly place themselves in these areas, even if they actually oﬀer solutions that cover several aspects of an iterative development process, wherein updates are made continuously. However, this confirms that this emerging theme is gaining ground.

With respect to the targets considered, component-based software and intensive software systems have been most investigated. These targets are characterized by the existence of different components, services, or subsystems. These can be independent of each other (e.g., micro services) or have strong dependencies and relationships amongst themselves and with the environment, as in many complex systems. However, in both cases, the presence of heterogeneous components makes it necessary to integrate these activities into continuous development and maintenance (e.g., DevOps). The component-based development paradigm is based on the concept of reuse within distinct components (e.g., services), enabling integration. Once integrated and implemented, these components must enter a continuous dimension, that is, to know and analyze the behavior after the integration and not only of the single component; therefore, they are also important in the context of performance.

Next, the results show the targets of cloud and distributed systems, characterized by the technology stack and infrastructure complexity. Cloud nodes may attain performance orders of magnitude worse than other nodes \cite{armbrust09}. For instance, if during the hosting of a mission-critical service or a scientific application, performance variability and availability become a concern, cloud monitoring is required to continuously measure and assess the infrastructure or application behavior (in terms of performance, reliability, power usage, and security) to adapt the system to changes or to apply corrective measures. Generally, we can observe that in open systems characterized by uncontrolled requests, continuous engineering, which supports their integration and evolution, is fundamental to identifying the occurrence of further problems not observed before.

Finally, we note that more recent and innovative targets are not yet widely investigated. For instance, in the case of data-intensive systems, the massive use of big data and machine learning requires efficient management of resources, performance, and security.

\smallskip
\noindent\fbox{
    \parbox{\linewidth}{
    \noindent
    \emph{Main findings:}
    \begin{itemize}[leftmargin=*]
        \item Significant attention is paid to continuous monitoring of data, which represents a fundamental capability to provide CSE;
        \item CSE and DevOps are gaining ground: most studies offer solutions covering several aspects of the iterative development process where updates are made continuously;
        \item Software intensive systems, especially when component-based, are the most investigated ones: their heterogeneous components require the integration of their activities in a continuous development and maintenance;
        \item More recent and innovative targets, such as cloud and data intensive systems, have not been widely investigated.
    \end{itemize}
    }
}

%% file: results-RQ2.tex
\section{What and how performance problems have been addressed (RQ$_2$)}
\label{sec:results:rq2}
\noindent
We classified the \SelectedFullPapers selected papers by considering the target problems addressed by them (we divided the target problems in five categories) and their research contributions (we identified seven diﬀerent types of contributions). In particular, the selected papers were thematically associated with at least one target problem and at least one research contribution based on their research directions and scope. We provide a detailed overview of the performance problems that have been addressed and illustrate them with  exemplary papers.

\subsection{Research problems}
\begin{figure}[htbp]
    \centering
    \includegraphics[width=.95\linewidth]{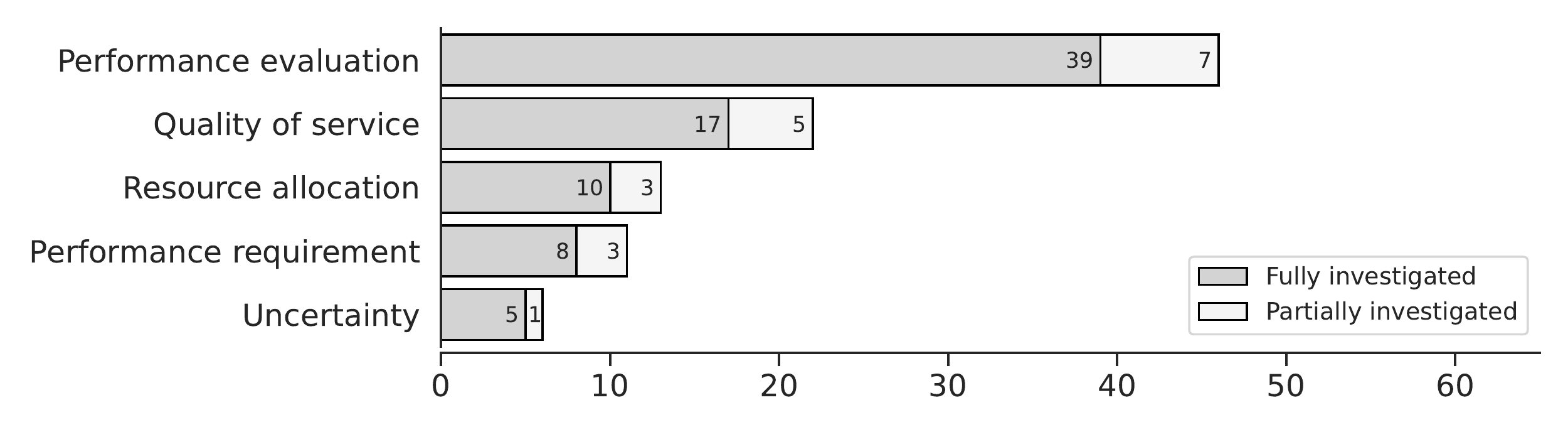}
    \caption{Research problems - results}
    \label{fig:target_problem}
\end{figure}

\noindent
Figure~\ref{fig:target_problem} presents the problems targeted by the selected papers (\underline{\textbf{target problem}}). 
The results obtained confirm that most of the selected papers ($46$ papers over \SelectedFullPapers) are focused on \emph{performance evaluation}. They include methods and techniques aimed at evaluating or predicting the expected performance within a continuous (re-)engineering of the system. 
~For instance, \ref{bib:Voneva2020} presents an approach for continuous integration of performance model during agile development, by optimizing the learning of parametric dependencies; whereas \ref{bib:Incerto2017} presents an approach that allows continuous performance adaptation through model predictive control. 

We can see that $22$ selected papers aim to provide \emph{QoS}, where the target is to satisfy different the requirements/properties for the overall quality of the system, including guaranteeing a certain level of performance.
For instance, in \ref{bib:Walter2017}, the authors proposed an approach for the automated extraction of quality-aware architecture models to explore the performance properties in design-time and runtime scenarios. In \ref{bib:DeSanctis2020}, the authors presented a method to apply the DevOps paradigm in the context of QoS-aware adaptive applications.
In addition to performance, some of the selected papers provide specific support for other properties, such as availability (\ref{bib:Trubiani2017} and \ref{bib:Garlan2004}), reliability (\ref{bib:Calinescu2011}, \ref{bib:Epifani2009}, \ref{bib:Pitakrat2018} and \ref{bib:RodrigoCalheiros2011}), privacy (\ref{bib:Heinrich2016}), scalability and resilience (\ref{bib:Cholomskis2018}).

Moreover, $13$ of these considered \emph{resource allocation}. This is an expected outcome that is strictly related to our query. The problem of allocating resources
is a great challenge during continuous development and operation and can severely impact the fast/frequent delivery of the team. 
For instance, in \ref{bib:Brunnert2017}, resource profiles were used to detect (performance) changes in enterprise application versions and to support capacity planning for new deployment. In \ref{bib:Willnecker2018}, performance models were extracted from the continuous monitoring of operational data, and then used to simulate performance metrics (resource utilization, response times, and throughput) and runtime costs of distributed and component-based applications.

Furthermore, $11$ selected papers addressed the \emph{performance requirement}.
Recently, with the increasing complexity of systems, high-performance requirements have become inevitable. For instance, \ref{bib:Yasaweerasinghelage2019} proposed a tool to discover secure architectural design options satisfying cost and performance requirements, while \ref{bib:Trubiani2018} proposed a performance-driven software refactoring approach to continuously assess and act with the aim of satisfying performance requirements.
 
Notably, the mapping revealed that and \emph{uncertainty} is an emerging target problem ($6$ papers). 
These papers offered interesting solutions to evaluate and predict the performance of the system, even in uncertain conditions. These works considered different types of uncertainty that inevitably affect the accuracy of quality evaluations. External uncertainty stems from the complexity and unpredictability of the physical environment (as in \ref{bib:Bao2017} and \ref{bib:Trubiani2017}) and of the non-software-controlled parts (such as cloud networks \ref{bib:RodrigoCalheiros2011}). Other papers (\ref{bib:Gerostathopoulos2016} and \ref{bib:Aleti2018}) referred to both external and internal uncertainty, where the latter referred to the uncertainty stemming from the complexity of the software system itself (large codebases, multiple development teams, and diverse development practices and standards, etc.). A different characterization of the uncertainty has been described in \ref{bib:Trubiani2019}, distinguishing uncertainty in parameters, models, code, and monitored data and providing a solution to detect violations of performance requirements in the DevOps process.

\subsection{Research contributions}

\begin{figure}[htbp]
    \centering
    \includegraphics[width=.95\linewidth]{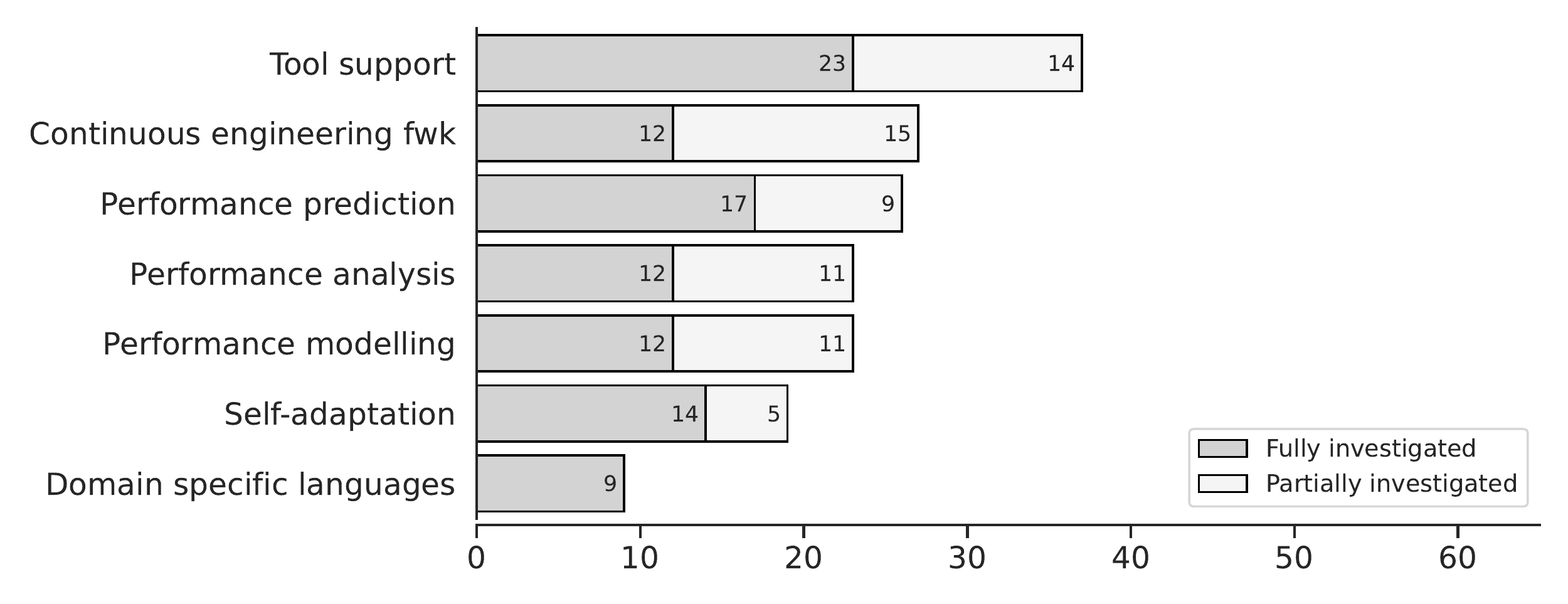}
    \caption{Research contributions - results}
    \label{fig:research_contribution}
\end{figure}

\noindent
Figure~\ref{fig:research_contribution} describes the principal \underline{\textbf{research contributions}}. From our analysis, we can state that most of the studies contribute to the continuous performance assessment of the system with \emph{performance analysis, performance modelling, and performance prediction} approaches. In particular, \emph{performance analysis} approaches, including methods and tools that allow the analysis of performance during the continuous engineering of the system, have been proposed in $23$ papers.
For instance, \ref{bib:Pooley2010} proposed a method for continuously assessing the performance requirements using the system architecture. In contrast, \ref{bib:Bernardi2018} proposed a performance analysis approach based on process mining.
In addition, $23$ papers proposed approaches that contributed in the process of modelling the performance of the system (\emph{performance modelling}). For instance, \ref{bib:Perez-Palacin2019} presented an UML profile for the design, quality assessment, and deployment of data-intensive applications, whereas \ref{bib:Li2017} generated layered queuing networks for the performance analysis of data-intensive applications.
Furthermore, $26$ papers have proposed approaches for \emph{performance prediction}. For instance, \ref{bib:Willnecker2019} proposed a model-based approach that was integrated into a continuous delivery pipeline, allowing to prediction of memory management and garbage collection behavior.  In contrast, \ref{bib:Incerto2017} proposed an approach of continuous assessment and adaptation through model predictive control.

In addition, we identified $19$ papers that considered \emph{self-adaptation}, allowing continuously running software systems to operate in changing contexts while meeting their quality requirements. 
Most of the approaches use performance evaluation to achieve the desired QoS objectives in software systems. Such approaches aims to provide, for instance,
re-configurable SAs (\ref{bib:Liu2007} and \ref{bib:VonMassow2011}),  self-adaptive microservice architecture (\ref{bib:Cholomskis2018}), or self-adaptation of performance parameters on the running system (\ref{bib:Incerto2016} and \ref{bib:Calinescu2011}). 
In contrast, \ref{bib:Ehlers2011b} presented a framework for self-adaptive monitoring to investigate the performance anomalies in component-based software systems. Notably, only \ref{bib:RodrigoCalheiros2011} partially addressed self-adaptation under uncertain conditions. In  particular, the authors modeled and analyzed the behavior and performance of Cloud systems to adaptively serve end-user requests, considering the uncertain behavior of resources and networks.

A new research direction is emerging that is related to \emph{domain specific languages}, for which we report only $9$ selected papers out of the \SelectedFullPapers considered. For example, \ref{bib:Bao2017} extended the hybrid architecture analysis and design language to support the modeling of environmental uncertainties and performed quantitative evaluations against various performance queries. In \ref{bib:Castellanos2019}, the authors proposed a domain specific model approach to design, deploy, and monitor the performance quality in applications related to big data analytics.
\ref{bib:Vögele2018} proposed a domain-specific language that allowed for the modeling of workload specifications of session-based systems for load testing and performance prediction.

A large number of papers ($37$) provides a \emph{tool support} to aid in performance engineering research. In addition, a considerable number of approaches have cosidered a \emph{continuous engineering framework} ($27$) . Both \emph{tool support} and \emph{continuous engineering framework} are generally considered together along with other research contributions. For instance, \ref{bib:Walter2017} provided a tool support for the extraction of architectural performance models based on monitoring of log files, whereas \ref{bib:Wert2015} proposed a continuous engineering framework to automatically build and parameterize performance models for large scale enterprise systems.

\subsection{Discussion}
\label{sec:discussion:rq2}
\noindent
Most studies in this area are aimed at solving performance evaluation and assessment problems. This is a relevant goal in the development of modern systems, where increasingly agile paradigms require that performance analyses be in a continuous dimension to be effective.

Notably, 22 papers were found to focus on QoS. Of them, 15 papers used a self-adaptation approach, implying that in 15 of the 18 papers, the self-adaptation approach targeted QoS rather than just performance. 

Managing the uncertainty of the system behavior is an emergent topic (papers addressing this issue are recent publications), and it is cross-cutting to the target problems previously mentioned. Although having the complete model of the system represents the ideal situation, in practice, only partial and limited measures are available. Consequently, specialized performance analysis or prediction techniques must work with uncertain knowledge. The proposed studies and their discussions on the diﬀerent types of uncertainty highlight relevant issues and oﬀer new research ideas.

More than half of the target problems considered were addressed using the support of tools and were well partitioned between performance prediction, performance analysis, and performance modeling. The other target problems were dedicated to self-adaptation approaches and, to a lesser extent, domain specific languages. The latter result shows that few studies have exploited abstraction for continuous performance control, although they consider large heterogeneous runtime data. Raising the level of abstraction of the specification would favor increased automation and interoperability.

Finally, in most studies (52 of 63 papers), support for CSE is provided by means of dedicated tools or frameworks. This is an interesting result, as in the context of this study, the quality and performance requirements demand the support of continuous engineering frameworks or dedicated tools.

\smallskip
\noindent\fbox{%
    \parbox{\linewidth}{%
    \noindent
    \emph{Main findings:}
    \begin{itemize}[leftmargin=*]
        \item Performance prediction, performance analysis and performance modelling have been further explored in order to offer adequate support in continuous developing;
        \item A relevant number of self-adaptation approaches have been proposed to ensure the quality of services (including performance); 
        \item Quality and performance requirements demand for the support of continuous engineering frameworks or dedicated tools.
    \end{itemize}
    }%
}

%% file: results-RQ3.tex
\section{What instruments have been adopted (RQ$_3$)}
\label{sec:results:rq3}
\noindent
Performance analysis can be conducted by adopting multiple techniques with different output-targeted metrics and with the support of different types of input data. In this section, we aim to identify the instruments that are adopted more often in the context of the study.

\subsection{Input data}
\noindent
Figure~\ref{fig:rq3}(a) reports the input \underline{\textbf{data}}. In regard to CSE, a system is continuously monitored to feed performance indices back into a performance model that supports predictive analyses. 

A total of $61$ and $32$ papers (of \SelectedFullPapers) used \emph{runtime/monitored} and \emph{performance model} as input data, respectively. Lesser number of papers consider the performance model as the input data because performance models have only been integrated in the last few years in software engineering processes.
Whereas runtime performance assessment and fixing have long been considered as a common practices.%

Majority of the papers ($34$) that used monitored data also proposed a continuous approach to monitoring performance features and then used them mainly for analysis and prediction (e.g., \ref{bib:Ehlers2011b}, \ref{bib:Mazkatli2020}, \ref{bib:Cholomskis2018}, and \ref{bib:Keck2016}). In contrast, other studies applied simulation (\ref{bib:Bao2017}), system execution modelling \cite{SEM} (\ref{bib:Falkner2018}), and  performance model generator leveraging execution data (e.g., \ref{bib:Willnecker2019}).
However, papers that used performance model mainly adopted UML + MARTE (e.g., \ref{bib:Bernardi2018} and \ref{bib:Bao2017}),  queuing network (\ref{bib:Pooley2010}, \ref{bib:Li2017} and \ref{bib:Trubiani2017}) or the
Palladio component model (\ref{bib:Vögele2018} and \ref{bib:Willnecker2019}). 

On the $61$ papers using runtime/monitored data, only $five$ papers did not provide explicit information about it, while employing performance models. In particular, \ref{bib:Pooley2010} considered changing the non-functional requirements and  provided a method for continuously assessing the performance requirements using the system architecture. \ref{bib:Yasaweerasinghelage2019} proposed an approach for optimizing the performance, cost, and security in architectural design and it performed static analysis (i.e., using Taint analysis) for the identification of architecture-level vulnerabilities. \ref{bib:Trubiani2017} analyzed different and correlated QoS models, thereby reducing the overall uncertainty while continuously re-architecting the system. \ref{bib:Chatley2019} proposed the use of performance unit testing to explore the performance characteristics and continuously detect potential performance problems  throughout the development of a software system. \ref{bib:Chiprianov2014} defined an approach to predict system performance based on the event-driven architecture modelling of the system interactions and impacts. 

A significant number of papers ($14$) considered the \emph{software model} as an input to the process. This is likely due to the different notations that are usually adopted for representing software models, preventing an automated full integration in the performance assessment task. The papers that considered \emph{software model} as input fully investigated this aspect in most cases; some of these studies (\ref{bib:Heinrich2020}) proposed automated techniques based on model transformations to develop \emph{software model} as a first-class artifact in the software engineering process. 
The results here also provide evidence that \emph{data analytics} have not yet been largely considered in this domain ($9$ papers, including ~\ref{bib:Castellanos2019}, \ref{bib:Cholomskis2018}, and \ref{bib:Birngruber2015}). However, interest in data analytics has grown over time; thus, data analytics is expected to become a primary source of inputs in the next few years. For each input data point, 
the relationship between \emph{fully} and \emph{partially investigated} papers
is proportional (a good percentage of the methodologies/techniques are fully investigated), with the exception of \emph{software model} mentioned above, and \emph{data analytics}, for which the relatively high number of \emph{partially investigated} ones ($five$ over $nine$) is very likely due to the recent progress in the development of techniques for performance data analytics. 

\subsection{Methodologies and techniques}
\noindent
Figure~\ref{fig:rq3}(b) presents the \underline{\textbf{methodologies/techniques}} used in the selected papers. As expected, teìhe majority of papers are focused on with \emph{performance modeling} ($37$ over \SelectedFullPapers) and \emph{performance analysis} ($32$ over \SelectedFullPapers). It is necessary to build and analyze models to address the performance issues early in the lifecycle, as confirmed by the fact that many of these papers intend to address this aspect (~\ref{bib:Brebner2016},\ref{bib:Trubiani2019}, and \ref{bib:Spinnner2019}). \emph{Model based software engineering} and \emph{model driven engineering} techniques were also widely considered ($26$ papers), as they did not restrict the adoption of models to the performance domain. However, in several cases, models were also considered (as first-class citizens) in the software engineering domain (\ref{bib:Voneva2020}, \ref{bib:Arcelli2019}, and \ref{bib:VonMassow2011}). A considerable number of papers deal with \emph{performance model extraction} and \emph{performance testing} techniques (both $23$ papers)  which were typically adopted when studying performance issues on existing running software systems (\ref{bib:Vögele2018} and \ref{bib:Willnecker2016}). Finally, it is observed that although in the last few years the adoption of \emph{machine learning} and \emph{multi-objective optimization} techniques has spread in diverse fields in the context of CSE, they are still marginally considered, as $six$ papers each focus on these techniques. 
For each methodology/technique, the relationship between the \emph{fully-} and \emph{partially-investigated}
is proportional (i.e., a good percentage of methodologies/techniques are fully investigated). Only in regards to \emph{performance analysis} the number of papers that partially investigates this methodology increases,as is the case with \ref{bib:Mazkatli2020} and \ref{bib:Incerto2017}, which deal with Palladio component model and queueing networks.
It is observed that this occurs despite the fact that \emph{performance modeling} is fully investigated. In certain contexts, extensive performance analysis can be difficult owing to the lack of system measurements and parameter values. Hence, in such cases, performance modeling can be fully investigated, but the analysis remains marginal among the contributions of the papers. 

\subsection{Output measures and indices}
\begin{figure}[htbp]
    \centering
    \includegraphics[width=.95\linewidth]{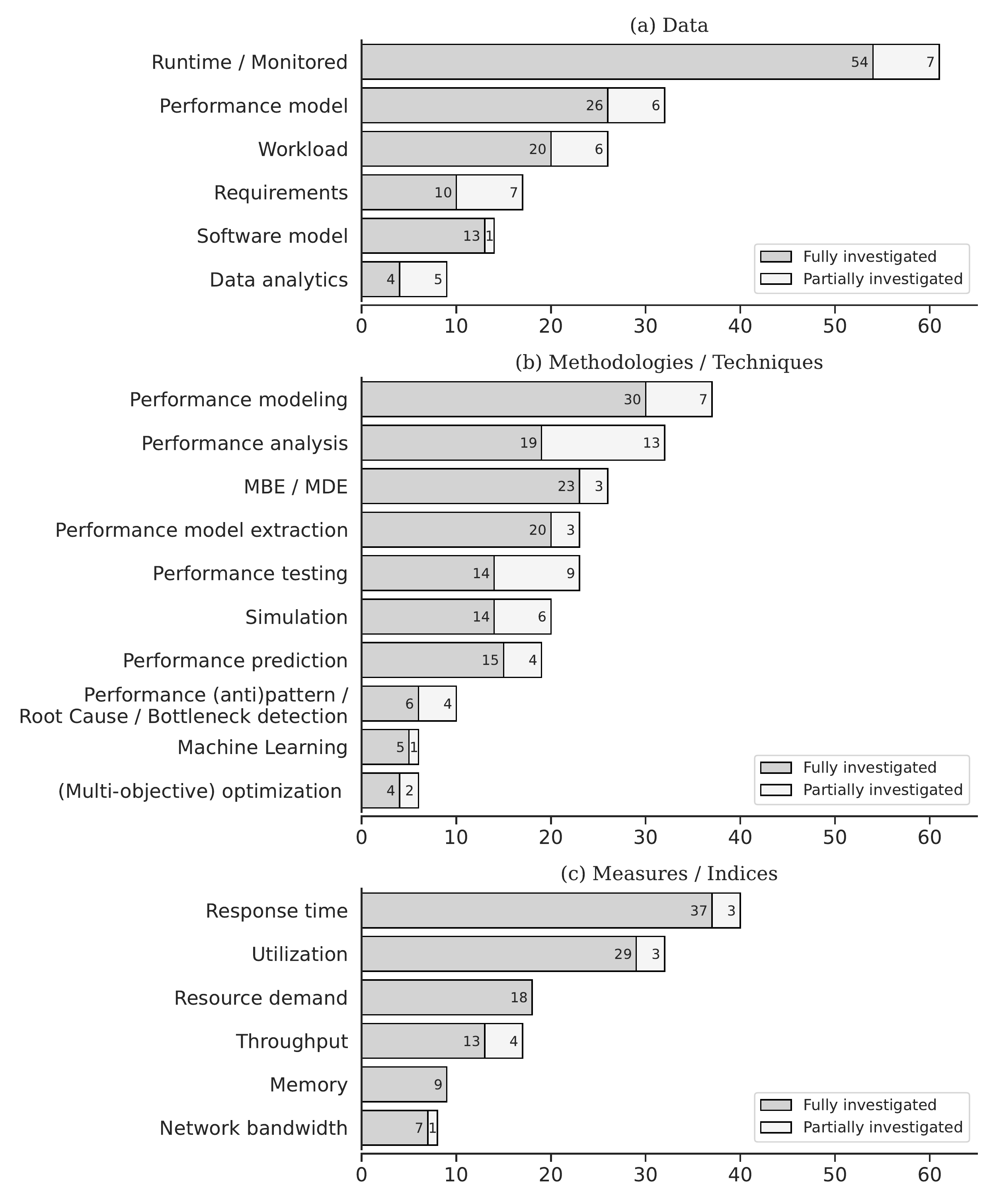}
    \caption{Data, methodologies/techniques, measures/indices - results}
    \label{fig:rq3}
\end{figure}

\noindent
Figure~\ref{fig:rq3}(c) shows the targeted output \underline{\textbf{measures/indices}}. The three typical performance indices, namely, \emph{response time} ($40$), \emph{utilization}, ($32$)  and \emph{throughput} ($17$), are the most widely targeted ones in the considered papers. Although, on one hand, this can be seen as an expected outcome, on the other hand it is somehow surprising that \emph{memory} ($9$) and \emph{network bandwidth} ($8$) have been significantly less studied in this context. These two measures may play crucial roles in the performance assessment of modern heterogeneous distributed software systems. Hence this result evidences a lack of investigation in this direction. 
These outputs are \emph{fully investigated}. Hence, no realistic consideration can be made on the few \emph{partially-investigated} ones. A few papers have considered multiple types of measures (\ref{bib:Aleti2018}, \ref{bib:Willnecker2016}, \ref{bib:Brunnert2017}, and \ref{bib:Calinescu2011}).

\subsection{Discussion}
\label{sec:discussion:rq3}
\noindent
We have analyzed the selected papers with the consideration that an approach can be categorized into three elements:  \emph{input data},  \emph{methodology/technique}, and \emph{output measures/indices}. Not all selected papers clearly undergo this schema, but the results obtained provide some interesting insights into the combination frequencies of these three elements. 

Based on a straightforward observation of the results, it can be seen that \emph{performance modeling} and \emph{analysis} techniques that takes as input \emph{runtime/monitored} data and produce \emph{response time} and \emph{utilization} indices as output have received the most consideration (till date). This is a relevant confirmation of what is expected performance analysts must do in the context of CSE processes. However, it is unexpected that regardless of the technique adopted,\emph{requirements} and \emph{data analytics} rarely enter the process to target \emph{memory} and \emph{network bandwidth}. This highlights a lack of investigation, which is crucial, especially in the domain of distributed heterogeneous software systems (CPS, edge computing, IoTT, and data-intensive systems). On the one hand, data analytics are ever more available and performance requirements are ever more stringent; on another hand, traditional performance measures (such as response time and utilization) in isolation do not provide an integrated vision of the system performance behavior, which can suffer from performance degradation owing to a bad usage of memory and poor network connection.

\smallskip
\noindent\fbox{%
    \parbox{\linewidth}{%
    \noindent
    \emph{Main findings:}
    \begin{itemize}[leftmargin=*]
        \item Approaches of performance modeling and analysis techniques that take as input monitored data and produce response time and utilization indices as output are widely used methodologies. Requirements and data analytics rarely enter the process to target memory and network bandwidth;
        \item Even if machine learning and multi-objective optimization techniques are being increasingly studied, they are still marginally considered in the context of CSE;
        \item Model-based  and model-driven techniques are widely considered, as they do not restrict the adoption of models to the performance domain, but in several cases, software models are also considered.
    \end{itemize}
    }%
}

%% file: results-RQ4.tex
\section{Current research gaps and future directions (RQ$_4$)}
\label{sec:results:rq4}

\noindent
In this section, we aim at detecting potential research gaps by visualizing the number of papers that lie at the intersection of research areas and target systems for each keyword of interest. In the following section, we discuss our findings in the categories of keywords:  \emph{target problems}, \emph{research contributions}, \emph{used methodologies and techniques}, \emph{used performance measures and indices}, and \emph{input data}. To represents our results, we developed bubble plots, as described in Section \ref{sec:method}. Moreover, we discuss the implications for future research based on the research gaps analyzed and future directions described in the selected papers.

\subsection{Target problems}\label{sec:results:rq4:target}
\begin{figure}[htbp]
\centering
\includegraphics[width=.9\linewidth]{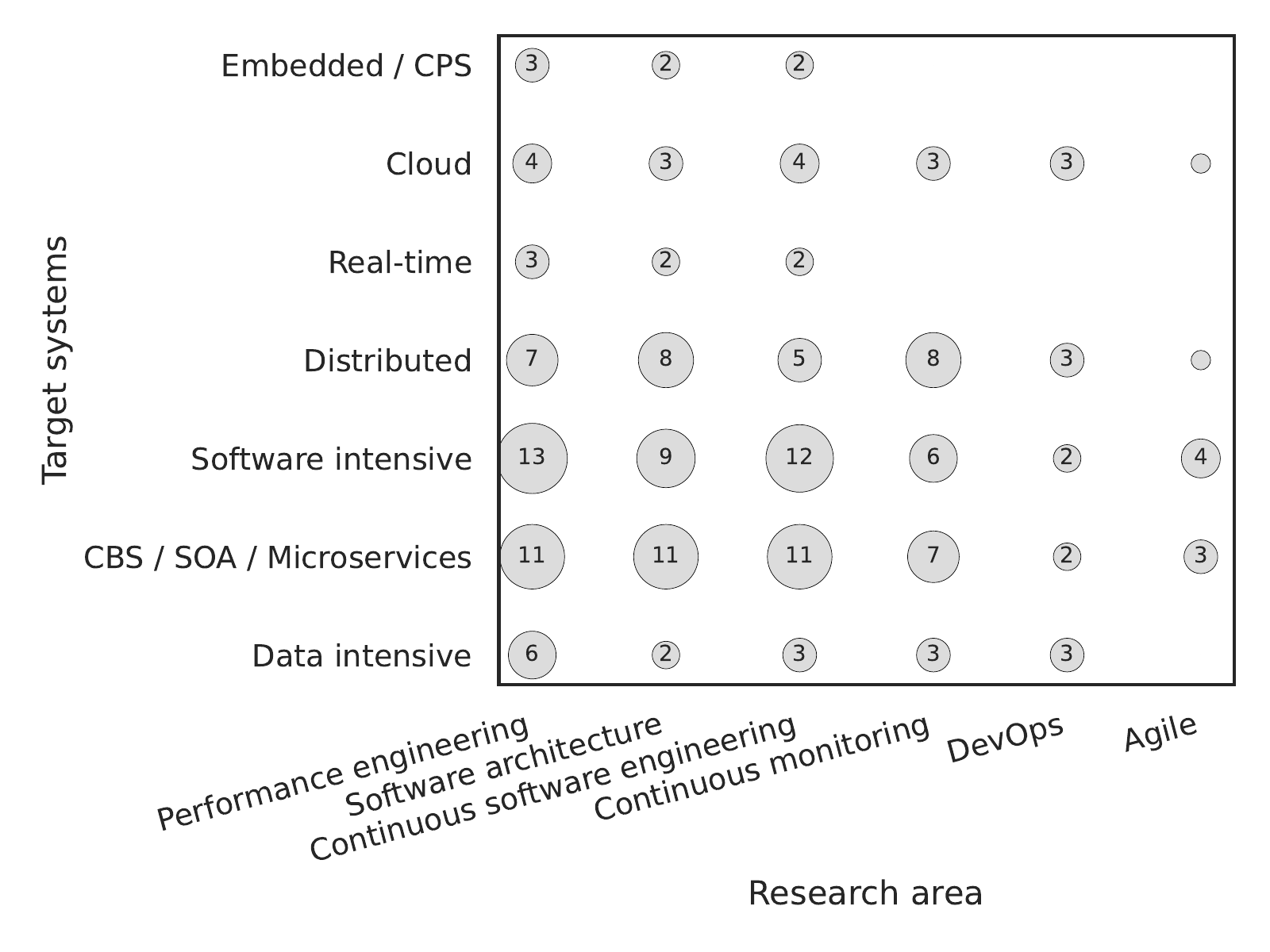}
\caption{Number of papers investigating the keyword \emph{performance evaluation (target problem)} at the intersection of research areas and target systems.}
\label{fig:bubble_target_problem_performance_evaluation__assessment}
\end{figure}

\begin{figure}[htbp]
\centering
\includegraphics[width=.9\linewidth]{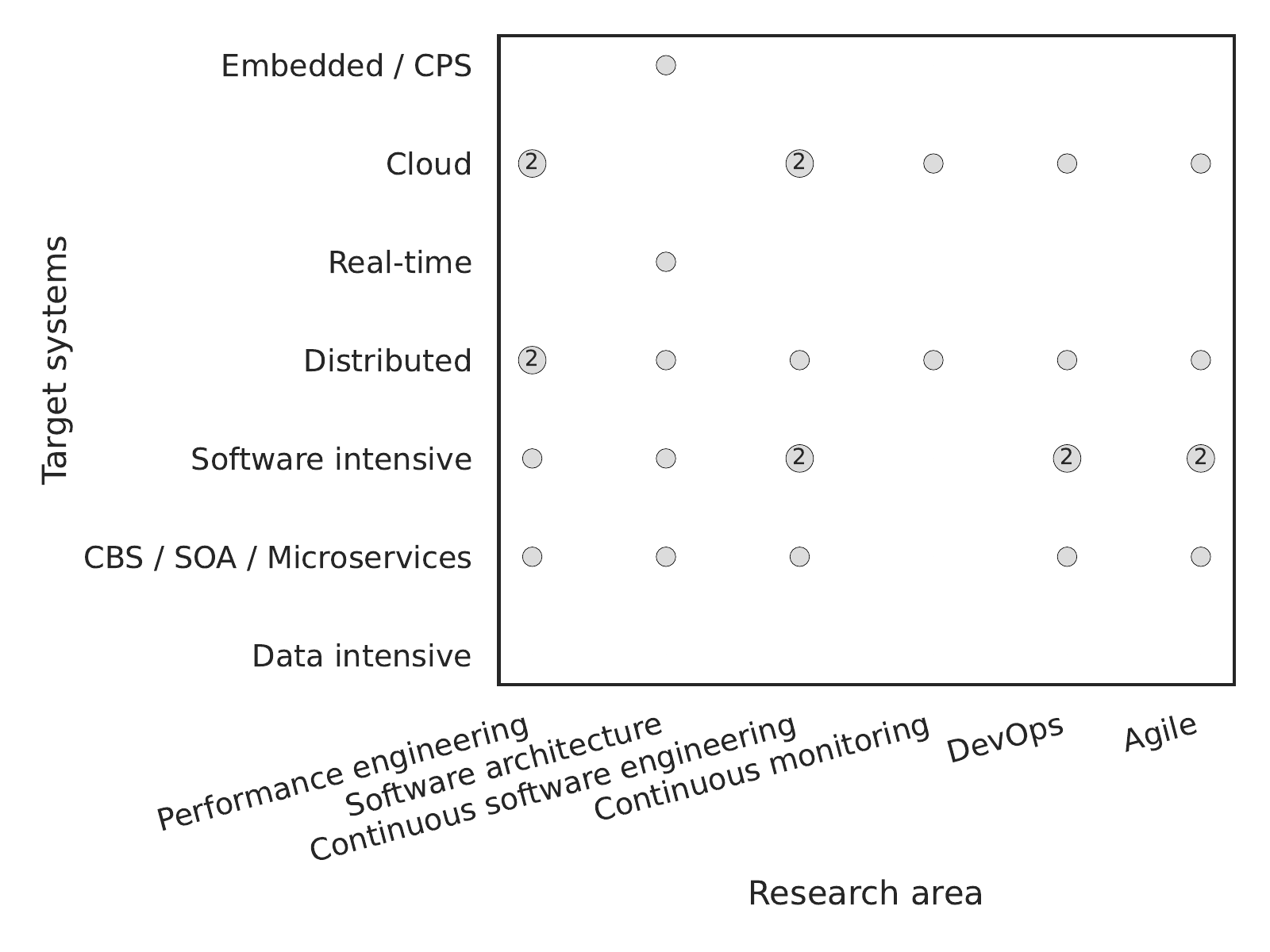}
\caption{Number of papers investigating the keyword \emph{uncertainty (target problem)} at the intersection of research areas and target systems.}
\label{fig:bubble_target_problem_uncertainty}
\end{figure}

\noindent
Figures \ref{fig:bubble_target_problem_performance_evaluation__assessment} and \ref{fig:bubble_target_problem_uncertainty} show the bubble plots for the \emph{Performance evaluation} and \emph{Uncertainty} target problems, respectively.
The plots present a very different situation.
Many studies have targeted problem of performance evaluation, especially in certain areas, and few papers have considered the uncertainty in general. A reason for this may be attributed to the fact that uncertainty has emerged only recently as a distinct concern in software engineering and its inclusion in continuous engineering practices is still very limited.

In contrast, because the general problem of performance evaluation is specifically targeted by our study, a greater number of papers are expected to consider it from Figure~\ref{fig:bubble_target_problem_performance_evaluation__assessment}, it is evident that certain research areas (continuous monitoring, DevOps, and agile) never intersect with certain target systems (real-time, embedded and CPS) when pursuing performance evaluation.
This could simply be due to the scarce adoption of DevOps and agile practices in these systems. 

A further gap appears in the area of the development of software intensive systems, even if it is continuously evolving  owing to the adoption of  new technologies, such as cloud computing, IoT, and artificial intelligence; Continuous engineering and DevOps can benefit from new performance engineering solutions to achieve more pervasive software (for example in smart cities, smart manufacturing, and smart mobility).

\subsection{Research contributions}
\begin{figure}[htbp]
\centering
\includegraphics[width=.9\linewidth]{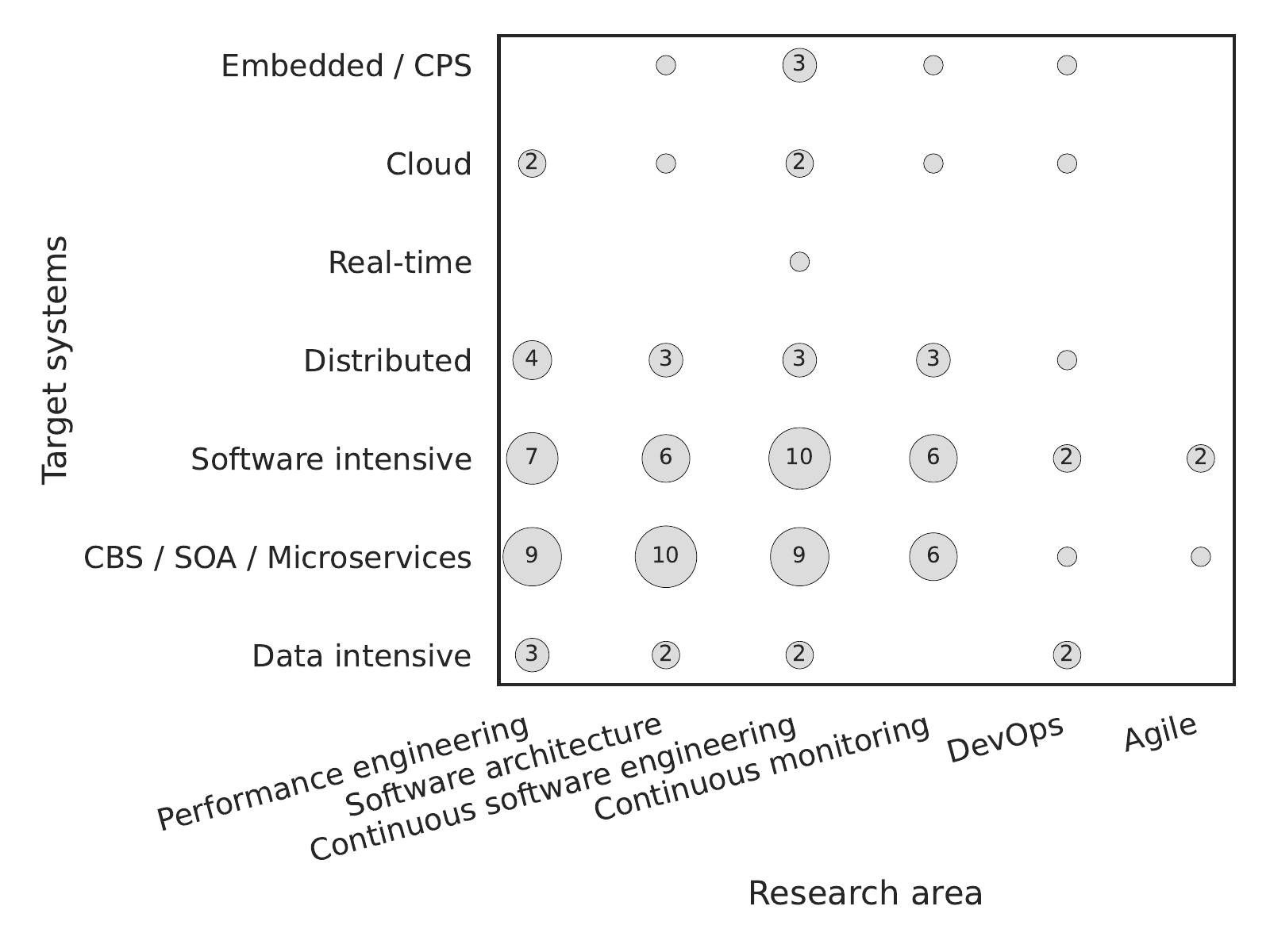}
\caption{Number of papers investigating the keyword \emph{continuous engineering framework (research contribution)} at the intersection of research areas and target systems.}
\label{fig:bubble_research_contribution_continuous_engineering_framework}
\end{figure}

\begin{figure}[htbp]
\centering
\includegraphics[width=.9\linewidth]{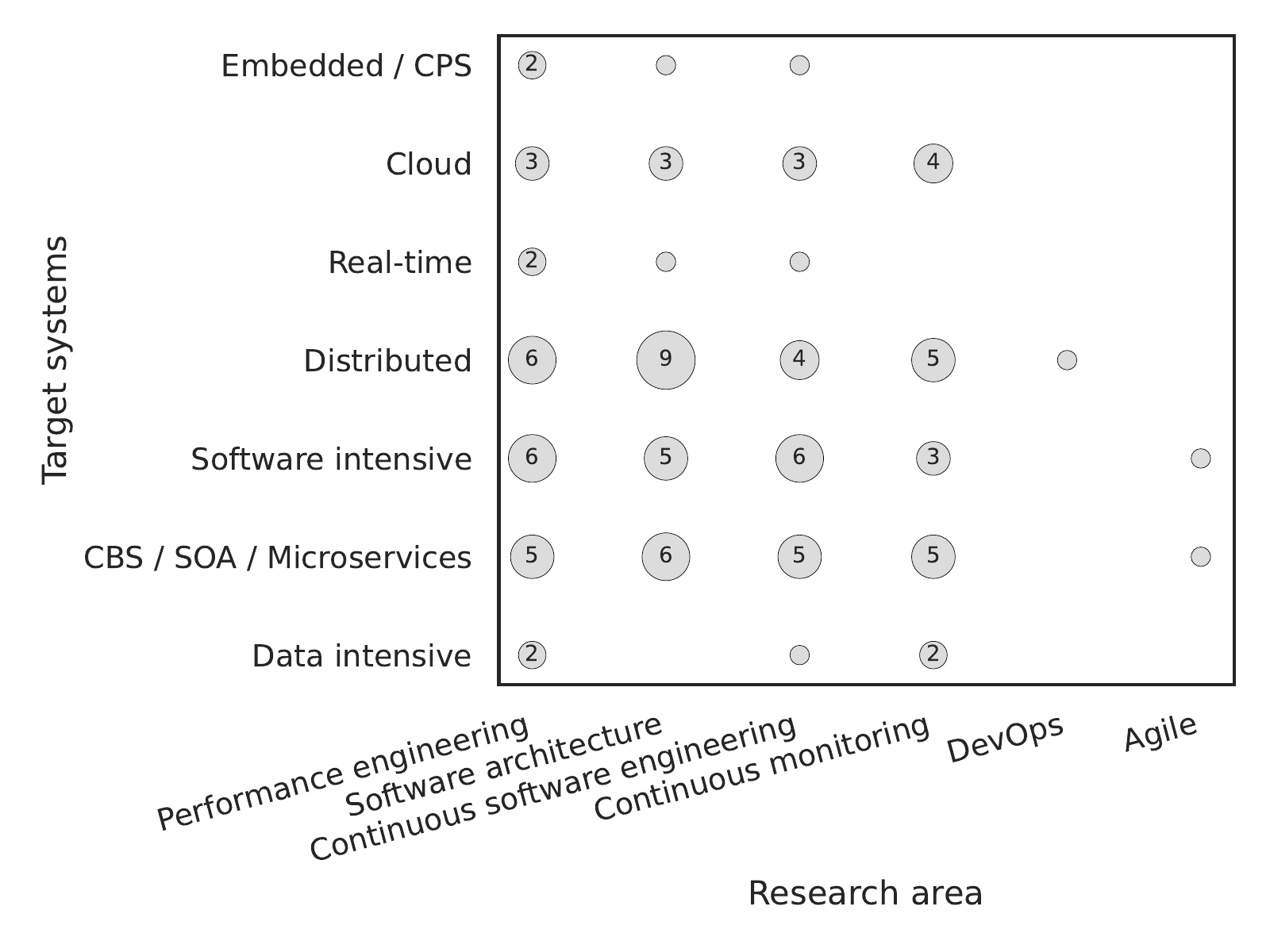}
\caption{Number of papers investigating the keyword \emph{performance prediction (research contribution)} at the intersection of research areas and target systems.}
\label{fig:bubble_research_contribution_performance_prediction_approach}
\end{figure}

\noindent
Among several research contributions identified, we reported the plots for \emph{continuous engineering framework} (Figure~\ref{fig:bubble_research_contribution_continuous_engineering_framework}) and \emph{performance prediction} (Figure~\ref{fig:bubble_research_contribution_performance_prediction_approach}) as we considered them to be relevant for the purposes of our study.
In Figure~\ref{fig:bubble_research_contribution_continuous_engineering_framework}, we observe that most of the papers proposing a novel continuous engineering framework are gathered in the lower half of the plot.
The target systems for which most of these frameworks are designed are CBS/SOA/Microservices, software intensive systems, and distributed systems.
Predictably, CSE is the most targeted research area in this field.
However, continuous engineering frameworks are rarely or never proposed in real-time and embedded systems.

Figure~\ref{fig:bubble_research_contribution_performance_prediction_approach} shows that only a few papers proposed performance prediction approaches in embedded, real-time, and data-intensive systems.
In addition, only three papers appear to focus both DevOps and agile.
This may represent a research gap about to be filled in the next few years because approaches based on artificial intelligence and machine learning, such as those in the AIOps \cite{AIOPS} field, are rapidly emerging as a new way of modeling and predicting performance that can be more easily integrated with current DevOps practices.

\subsection{Input data}
\begin{figure}[htbp]
\centering
\includegraphics[width=.9\linewidth]{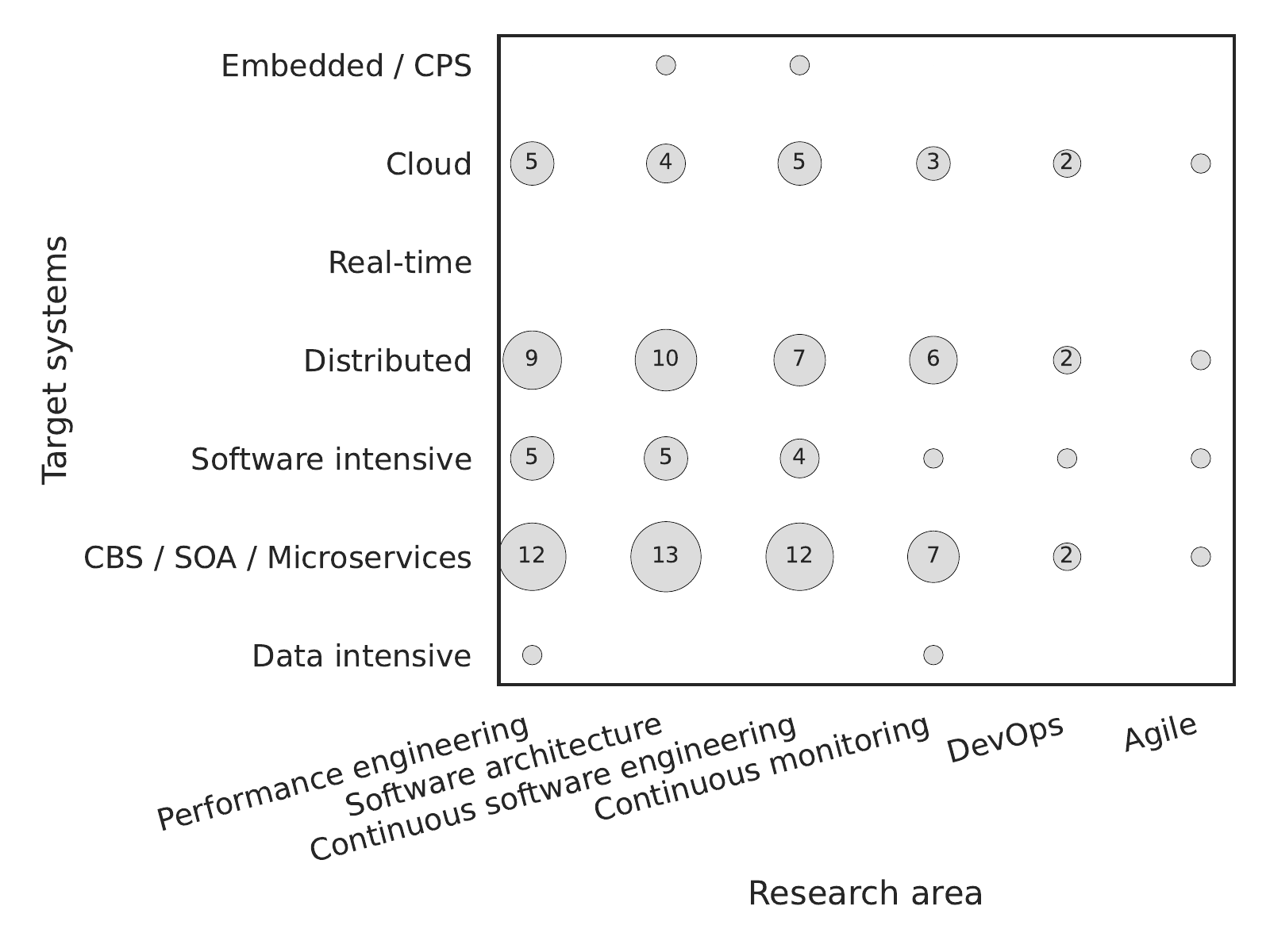}
\caption{Input data - Workload}
\label{fig:bubble_used_data_workload}
\end{figure}

\begin{figure}[htbp]
\centering
\includegraphics[width=.9\linewidth]{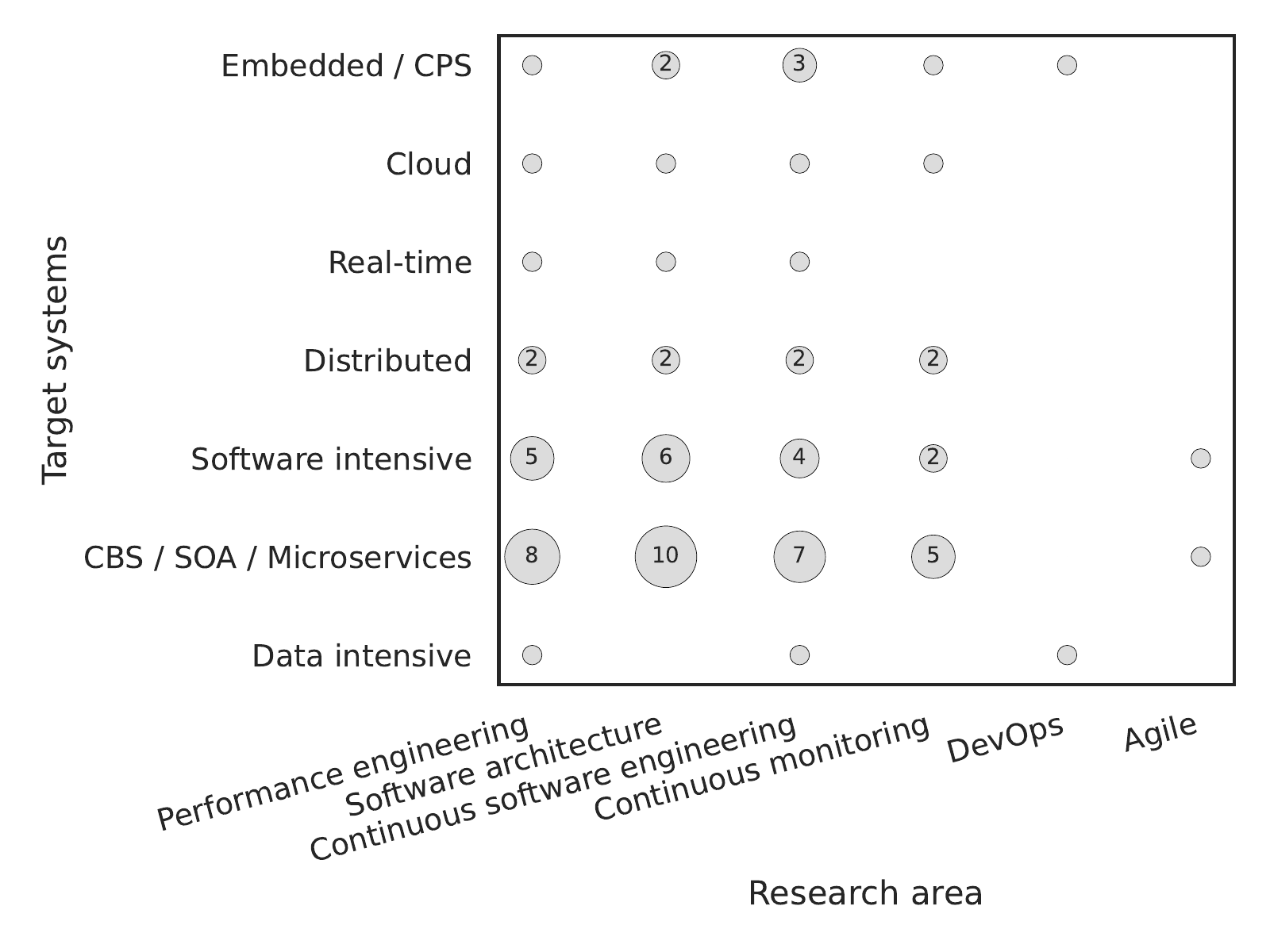}
\caption{Input data - Requirements}
\label{fig:bubble_used_data_requirements}
\end{figure}

\noindent
The types of data that are used as input to the approaches play a significant role in establishing the situations in which an approach can be applied and the type of information required to initiate the process. \emph{Workload} (Figure~\ref{fig:bubble_used_data_workload}) and \emph{requirements} (Figure~\ref{fig:bubble_used_data_requirements}) are the types of data, consideration of which appears to be related to the specific target system.
For instance, while the workload is often considered in the \emph{cloud}, \emph{distributed}, and \emph{CBS/SOA/microservices} systems, it is almost never considered in \emph{embedded}, \emph{real-time}, and \emph{data intensive} systems. The lack of consideration of workload in embedded and real-time systems is expected, whereas in data intensive systems it presents a research opportunity. When considering the use of requirements, we are presented with a different situation. From the number of papers in the bubble plot, it appears that \emph{DevOps} and \emph{agile} do not put much emphasis on the requirements when assessing the performance; this is unexpected because they consider the specification of requirements in their processes. In addition, the \emph{software intensive} and \emph{CBS/SOA/micorservices} systems often consider the requirements as the starting point for the development of performance engineering approaches.

\subsection{Methodologies and techniques}
\begin{figure}[htbp]
\centering
\includegraphics[width=.9\linewidth]{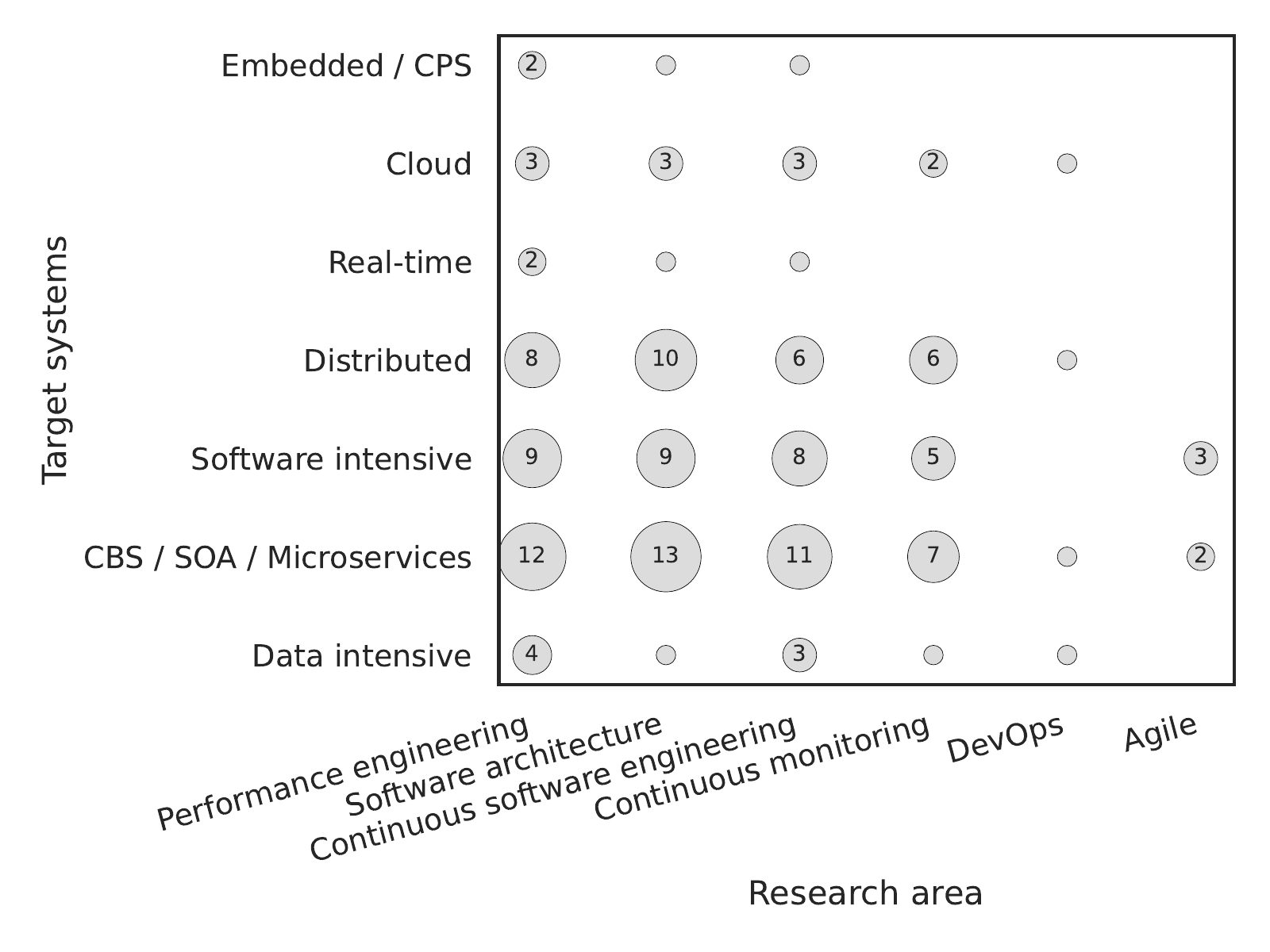}
\caption{Methodologies / techniques - Performance model}
\label{fig:bubble_used_methodologies__techniques_performance_model}
\end{figure}

\begin{figure}[htbp]
\centering
\includegraphics[width=.9\linewidth]{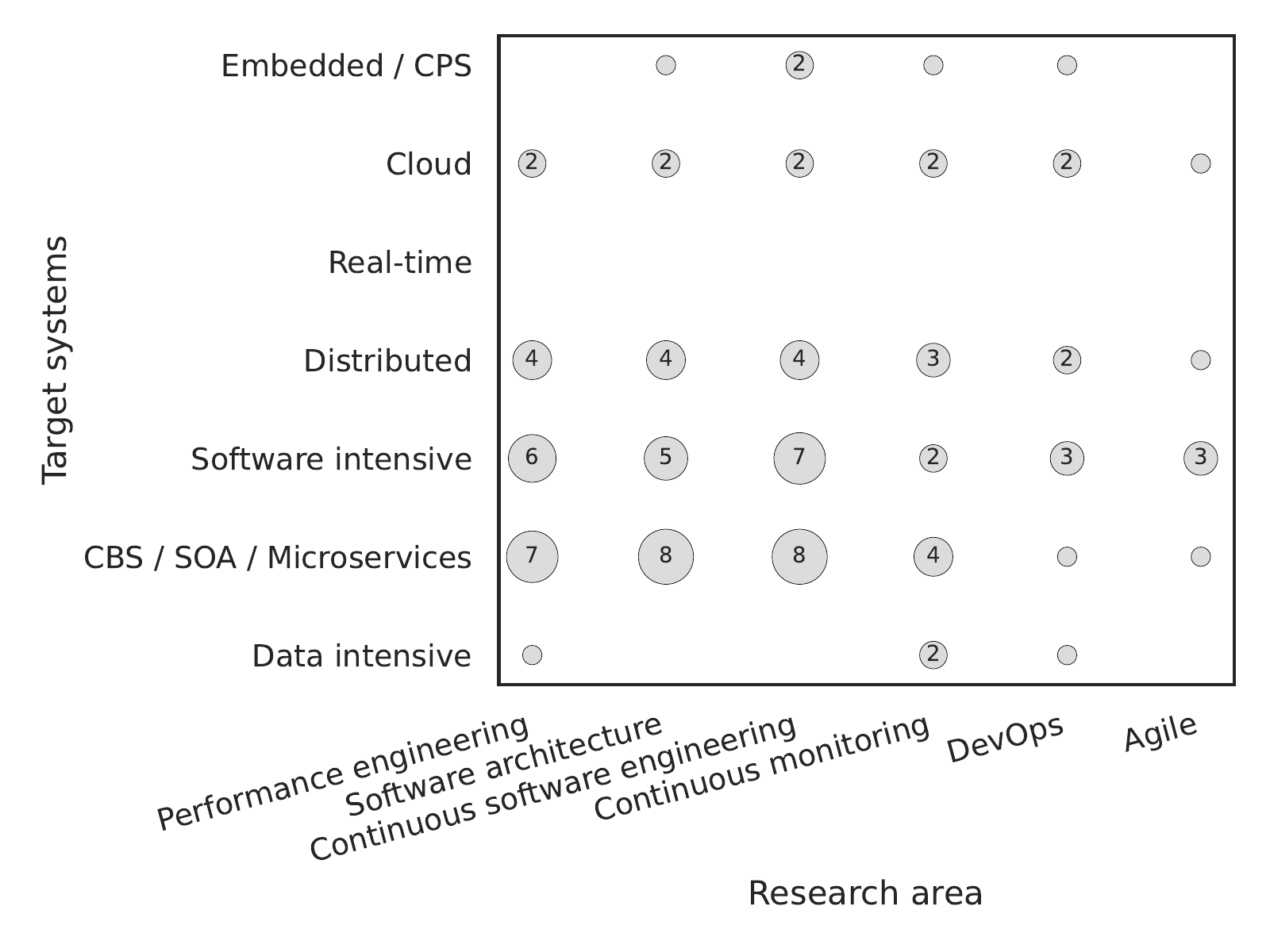}
\caption{Methodologies / techniques - Performance testing / Load Testing / Benchmarking}
\label{fig:bubble_used_methodologies__techniques_performance_testing__load_testing__benchmarking}
\end{figure}

\begin{figure}[htbp]
\centering
\includegraphics[width=.9\linewidth]{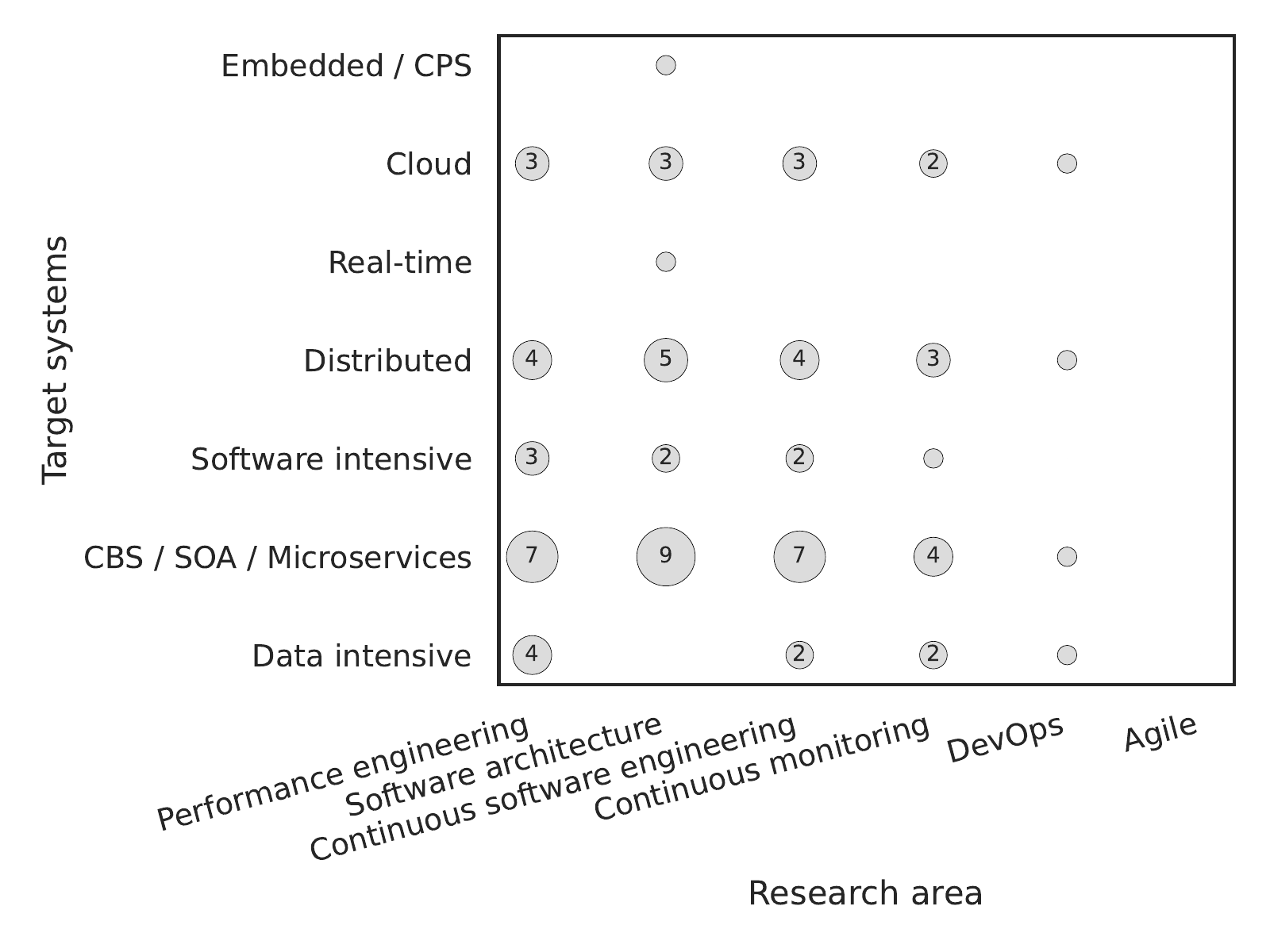}
\caption{Methodologies / techniques - Performance model generation / extraction}
\label{fig:bubble_used_methodologies__techniques_performance_model_generation__extraction}
\end{figure}

\begin{figure}[htbp]
\centering
\includegraphics[width=.9\linewidth]{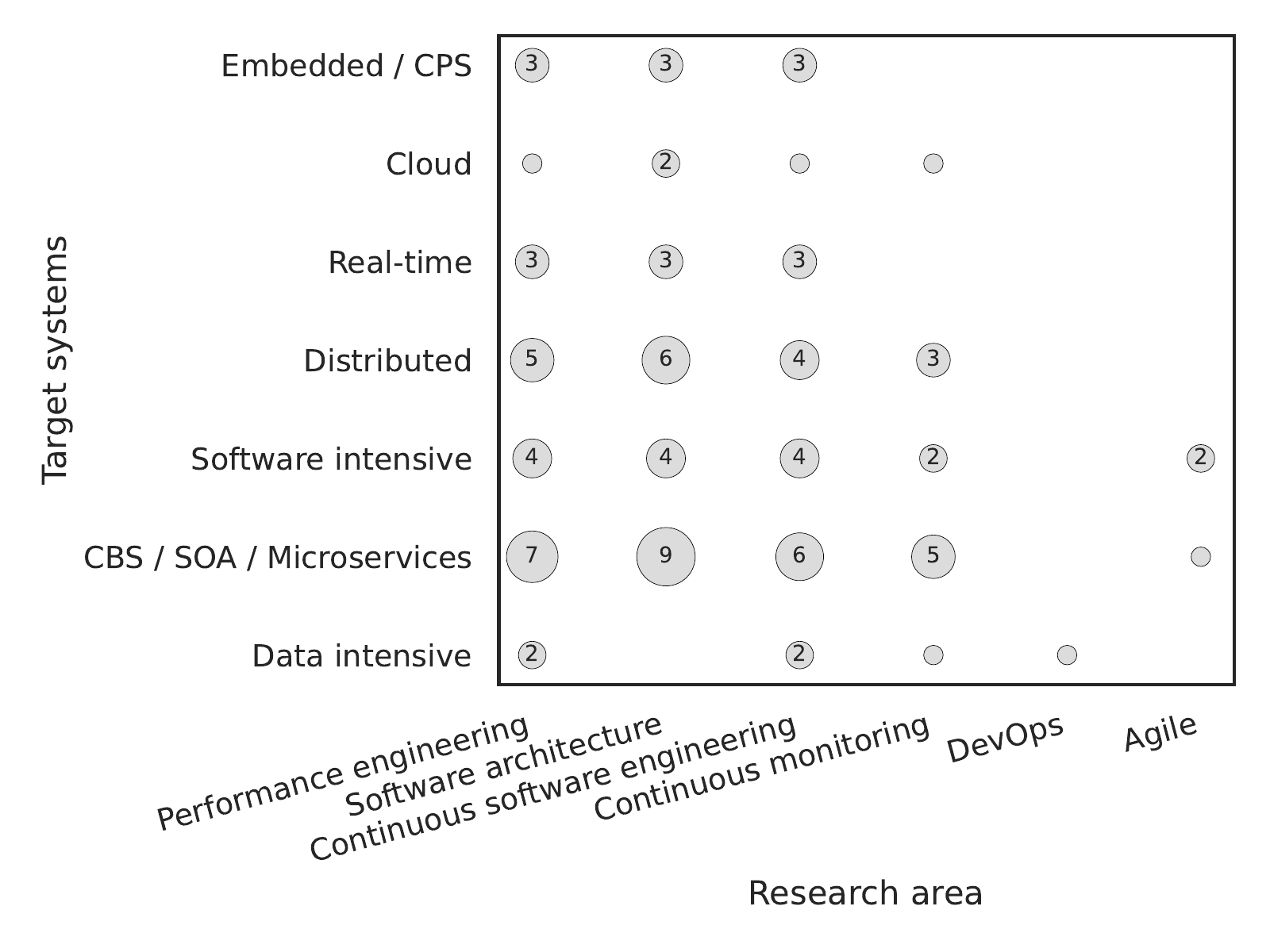}
\caption{Methodologies / techniques - Simulation}
\label{fig:bubble_used_methodologies__techniques_simulation}
\end{figure}

\noindent
The methodologies and techniques that are employed in the approaches of our study represent a compelling source of information for discovering the current research interests and gaps. For instance, when examining the use of performance models (Figure~\ref{fig:bubble_used_methodologies__techniques_performance_model}), it is clear that the DevOps and agile research areas lag behind the other areas in terms of the number of papers. A different picture is presented using performance testing, as shown in Figure~\ref{fig:bubble_used_methodologies__techniques_performance_testing__load_testing__benchmarking}. In this case, while all the research areas are almost equally represented, a lack of focus on the adoption of performance testing, load testing, and benchmarking is evident in the real-time and embedded systems, as the number of papers contributing to above mentioned aspects in the real time embedded systems are 0 and 3, respectively. Generation or extraction of a performance model (Figure~\ref{fig:bubble_used_methodologies__techniques_performance_model_generation__extraction}) is a special use case in the adoption of performance models. Therefore, unsurprisingly, we can still count a few papers in the DevOps and agile research areas, whereas distributed and microservice systems seem to rely the most on the automated generation of performance models.
Finally, in regard to the use of simulation (shown in Figure~\ref{fig:bubble_used_methodologies__techniques_simulation}), we observed that, in contrast to other methodologies, simulation has been employed in several papers for real-time and embedded systems. In addition, DevOps and agile appear to be less represented than in other research areas.

\subsection{Output measures and indices}\label{sec:sec:results:rq4:measures}
\begin{figure}[htbp]
\centering
\includegraphics[width=.9\linewidth]{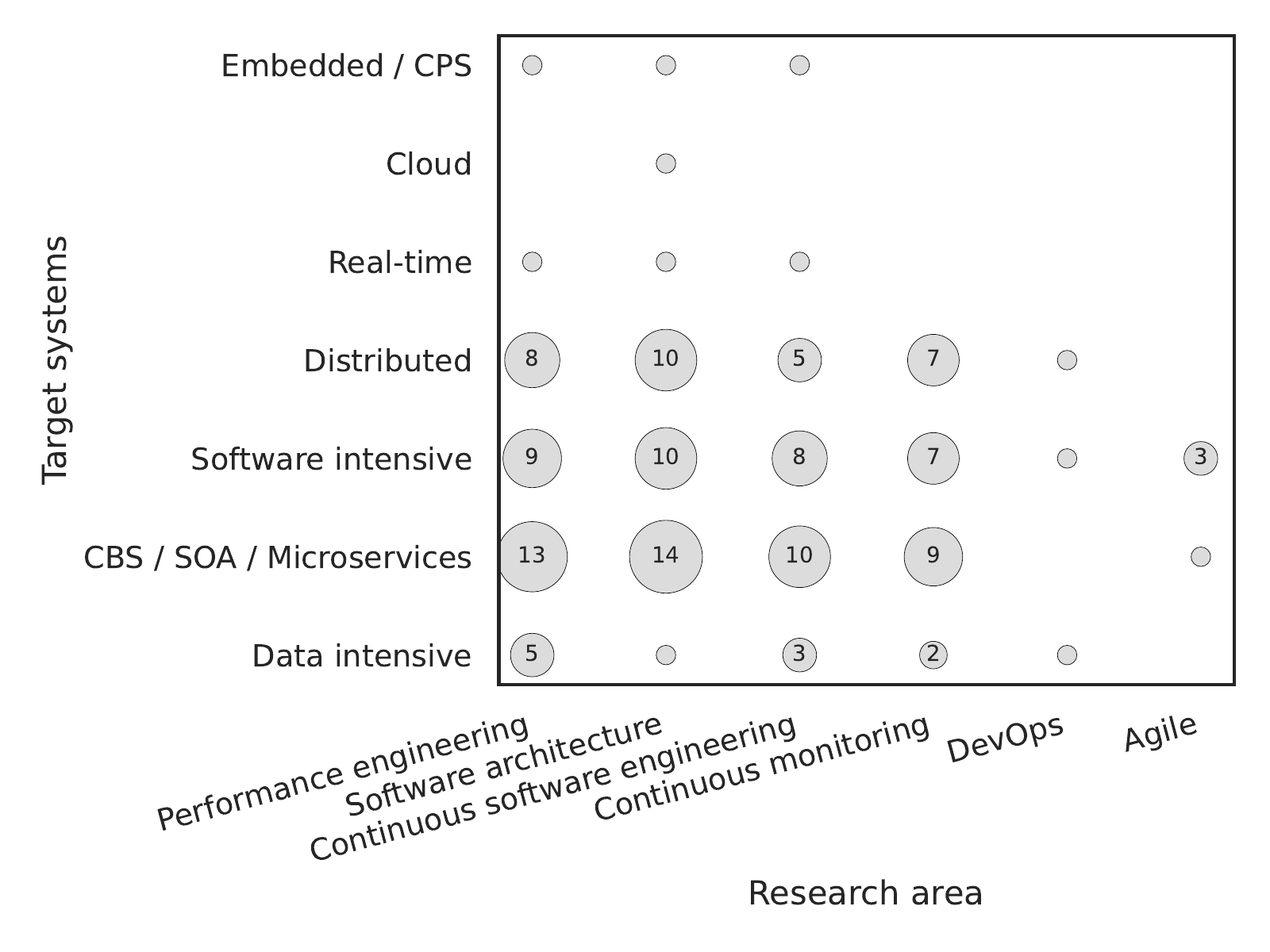}
\caption{Output measures / indices - Response time}
\label{fig:bubble_used_performance_measure__indices_response_time}
\end{figure}

\begin{figure}[htbp]
\centering
\includegraphics[width=.9\linewidth]{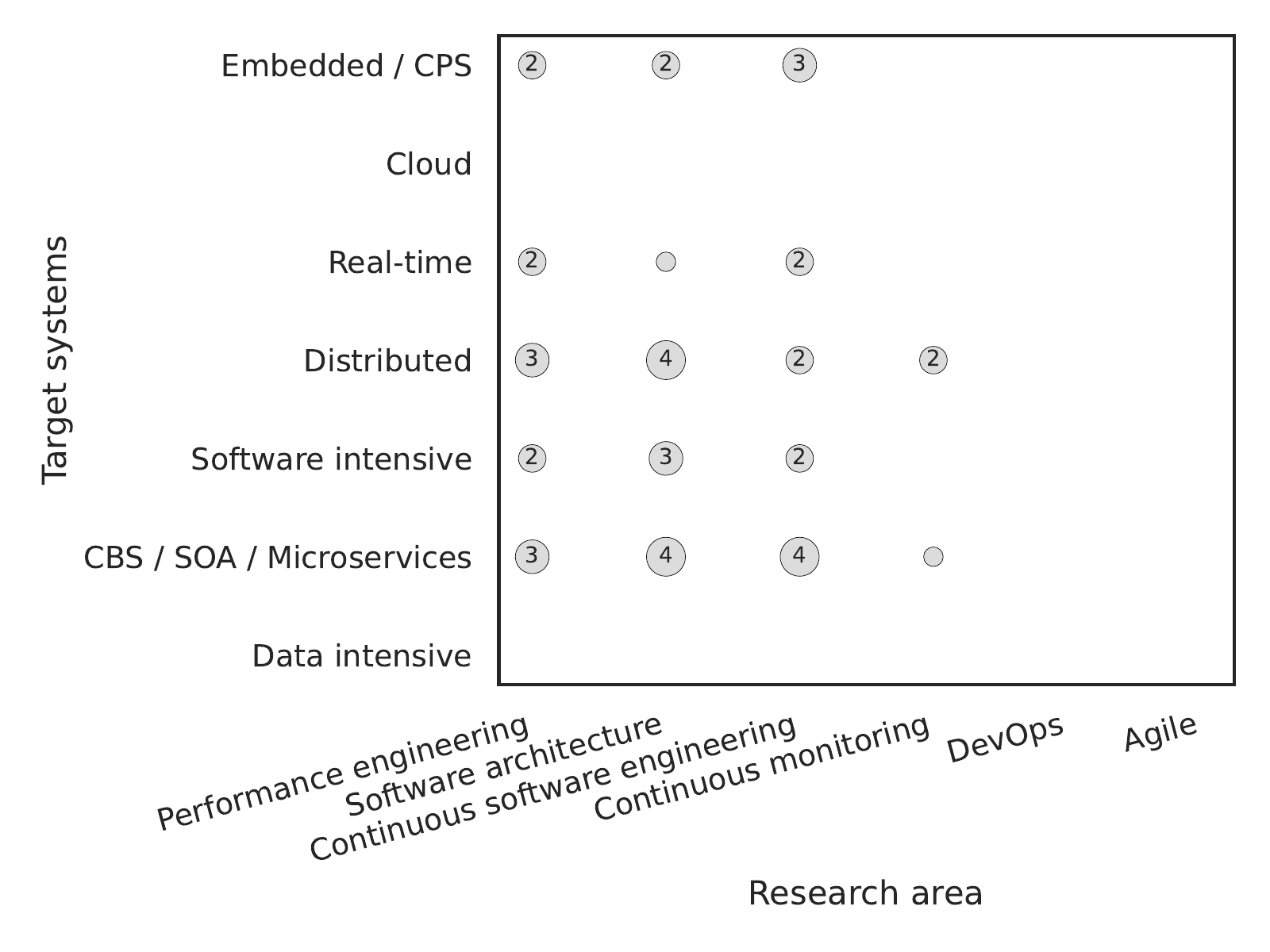}
\caption{Output measures / indices - Memory / Memory Leaks}
\label{fig:bubble_used_performance_measure__indices_memory__memory_leaks}
\end{figure}

\noindent
When investigating the performance measures and indices that were targeted as outputs in the selected papers, we observed two radically different situations in the number of papers that considered the keywords \emph{response time} (Figure~\ref{fig:bubble_used_performance_measure__indices_response_time}) and \emph{memory/memory leaks} (Figure~\ref{fig:bubble_used_performance_measure__indices_memory__memory_leaks}).
In software performance engineering, response time and memory are among the primary measures of interest as they are usually employed to assess the quality of service and operating costs, respectively. Nonetheless, memory is rarely considered in the papers selected for this study. Even more interesting is the fact that \emph{DevOps} and \emph{agile} seem to not consider memory at all in any target system. We expect memory use to become critical as ML and data-intensive systems continue to increase.
In contrast, response time is considered more often in general and in particular in the  domains of \emph{distributed systems}, \emph{software intensive systems}, and \emph{CBS/SOA/imcroservices}. 
This paints a picture in which performance measures that impact the quality of service are the foremost concern in research related to CSE, thus resulting in a substantial gap in the investigation of issues related to memory usage and how these can affect the cost of providing a service.

\subsection{Discussion on research gaps}
\label{sec:discussion:rq4}

\noindent
As is evident from the results reported previously in Section~\ref{sec:results:rq4}, there is a noticeable difference in the coverage between the more established and emerging research topics. A more general performance evaluation topic is a clear example. It is among the most investigated topics in our study, not only because it is understandably at the center of our inquiry, but also because it is a well-established research topic with a history spanning decades.

Nonetheless, when the same topic is combined with more recent trends in the software industry, such as agile and DevOps, it clearly lags behind with respect to the number of papers covering it. Apart from the obvious reason that more recent topics possibly receive less coverage in research, a more compelling justification can be found in the specific characteristics of agile processes and practices. In fact, agile cycles are usually considerably short on time as the main focus of such cycles is possibly the fast release of new features. While some tests, such as acceptance tests from the product owner, are understandably required before releasing a new feature, performance tests are usually not necessary as they are often very demanding in terms of the time and resources. A possible solution might be to classify new requirements with low and high performance priority and contemplate dedicating a considerable budget to performance analysis and testing.

Similarly, the specific needs of cloud-native systems do not always align with the requirements of performance evaluation. Most cloud-native systems adopt continuous tracing tools (Jaeger, OpenTelemetry, and AppDynamics) that record execution traces in highly distributed and dynamic environments. Such tools may allow performance analysis to be conducted as they usually provide additional information that is relevant to the performance, such as the execution time of remote calls. However, companies do not conduct performance analytics on distributed traces until performance does not compromise the user experience. Continuous assessment and improvement of performance are still not considered as prior quality tasks leading to software systems and services that quickly degrade their performance~\cite{AmellerFGMABCCF21}. This leads us to believe that a subsequent study that considers the industry (a multivocal literal review) may be of interest.

In contrast to agile, DevOps, and cloud, the topic of uncertainty in software engineering has been relevant for some time. Even so, it still represents an emerging topic in research, especially when considering performance evaluation in uncertain settings. Evaluating performance while maintaining uncertainty is inherently diﬃcult. The characteristics of software systems are becoming more dynamic owing to many factors: they are more distributed because they are moving on the cloud or are composed of third-party software, they are open to a wider group of users, and more importantly, they are expected to provide QoS guarantees in ever-changing environments. In this context, performance evaluation must consider fluctuating  workloads and great variability in resource demands, posing more challenges to the accuracy and validity of the prediction and analysis of performance results over time. To address this problem, researchers have started to study software performance under uncertainty, and our study finds that this is a research area not yet suﬃciently investigated, but that may possess great growing potential.

\smallskip
\noindent\fbox{%
    \parbox{\linewidth}{%
    \noindent
    \emph{Main findings:}
    \begin{itemize}[leftmargin=*]
        \item Although performance evaluation is a very well established research topic, the conjunction with the latest trends in the software industry, such as agile and DevOps, is clearly overdue.
        \item Researchers have started to study software performance under uncertainty. The research area has not yet been sufficiently investigated, but may possess a great growing potential.
        \item There is no evidence of the experience of companies in conducting performance analytics on cloud systems.
    \end{itemize}
    }%
}

%% file: future.tex
\subsection{Implication for future research}\label{sec:future}
\noindent
In this section we discuss the implications of this study and challenges for future research.

\subsubsection{Towards a culture of quality in CSE}
\noindent
The results obtained suggest an interest in a culture of quality, indicating that analysis and verification occur early in the CSE pipeline (as for testing in DevOps), making it easier to discover and fix defects with a collaborative approach to product improvement. In order to bring up the quality characteristics of software (architecture) and its continuous improvement (and re-architecting), QoS analysis must become an integrated activity in the entire lifecycle of software development lifecycle, which requires continuous exposure of quality characteristics exposed to analysis  (\ref{bib:Trubiani2017} and \ref{bib:Castellanos2019}). 

As discussed earlier, most of the selected studies focused on a few of the performance properties. However, there is a need to strengthen the support for various properties in both performance and, in general, software quality during the continuous engineering of the system. 
Some challenging quality aspects to consider are: verification of correctness properties, such as architectural mismatches (~\ref{bib:Castellanos2019}), 
evaluation of the architectural runtime models with respect to fidelity (\cite{BencomoGS19,bennaceur2014}) and usefulness for human inspection (\ref{bib:Heinrich2016}), extending the scalability by considering influencing factors such as variation in the complexity of user behavior in experiments \ref{bib:Heinrich2020}, supporting
state management and resource provisioning mechanism  \ref{bib:Junior2014}, and
introducing time consumption for memory allocation and release operations to increase the prediction quality of the model \ref{bib:Willnecker2019}.
Similarly, other QoS properties, such as the reliability~\ref{bib:VonMassow2011}, consistency ~\ref{bib:Calinescu2011}, safety and security, and availability~\ref{bib:RodrigoCalheiros2011} can be modeled and analyzed in a CSE framework.

\subsubsection{Performance engineering benefits in DevOps}
\noindent
DevOps is gaining widespread adoption in industry. However, its principles of rapid changes, development automation, and fast feedback loop (often relying on dynamic cloud environments) conflict with the complexity of the current performance engineering approaches~\cite{Bezemer2019}.
Thus, performance engineering frameworks should be improved for adoption in rapidly changing systems. The structures and behaviors of the modern systems changes frequently and requires continuous relearning of the failure models in order to retain the prediction quality \ref{bib:Pitakrat2018}. 

Moreover, these systems are characterized by a continuous stream of available data. Performance models should be built periodically, incrementally, or even continuously, and be triggered by changes to components in the monitored environment. Models can be quickly rebuilt once a potential problem is detected using only the most recent data only, and then used to compare with previous model results \ref{bib:Brebner2016}.

\subsubsection{Data-driven methods and Machine Learning}
Performance and load tests produce a large amount of data that can be difficult to analyze. Data-driven methods provide powerful insights into optimizing performance, building new features, and preventing problems with services, especially in distributed (enterprise) applications, in-memory databases, and big data systems \ref{bib:Heinrich2020}.

Despite developments  in modern software engineering technology, there is no established methodology for systematically employing performance engineering and data-driven engineering in continuous development.

In addition, performance evaluation based on machine learning can become an integral part of the continuous engineering process \ref{bib:Muller2019}. The performance model must learn autonomously and improve itself during system operation in a production environment \ref{bib:Grohmann2019}.

\subsubsection{Continuous controlling system uncertainty}
\noindent
As discussed in Sections \ref{sec:results:rq2} and \ref{sec:results:rq4}, uncertainty has been addressed marginally in the papers selected in this study. However, both in academia and industry, significant attention is paid to the contexts characterized by a high degree of uncertainty. To reduce uncertainty and obtain feedback on products/software as soon as possible, it is important to test assumptions and hypotheses in short cycles \cite{CSE}.

Continuous monitoring and frequent re-assessment and re-architecting are necessary to reduce the uncertainty in CSE and DevOps \ref{bib:Trubiani2019}. An efficient analysis method can be used to propagate the eﬀect of uncertain parameters in software systems and calculate the robustness of the performance indices, thereby enhancing the flexibility in addressing uncertainty (\ref{bib:Aleti2018}).

\subsubsection{Integration and abstraction}
\noindent
A recurring issue is the need to integrate methods and tools into the continuous and DevOps pipelines. Hence, it is necessary to design and develop performance engineering approaches that can be integrated with other tools and methods used in the context of this study \ref{bib:Spinner2016}. 

Continuous monitoring is a fundamental process in CSE. In the context of implications for future research, further investigation on adaptive monitoring and analysis infrastructure that can automatically update the system and performance/quality models is needed  (\ref{bib:Ehlers2011a}, \ref{bib:Ehlers2011b}, \ref{bib:Wert2015}, \ref{bib:Mazkatli2020} and \ref{bib:Calinescu2011}). 

External capabilities can be integrated using approaches to represents systems as black-box components by integrating black-box monitoring techniques \ref{bib:Calinescu2011}, or creating resource profiles describing specific enterprise applications (using standard measurement solutions), instead of relying on a custom solution to collect the required data \ref{bib:Brunnert2017}. This can be improved by supporting the collection of arbitrary information on the status of the monitored application, which requires the system integration of the corresponding type on a dynamic basis \ref{bib:ReinerJung2013}.

Using a higher abstraction level can help reduce the integration efforts. Future research can target the development of a model-based framework that considers the definition of (domain-specific) languages and automation mechanisms to ensure, by design, the potential for the monitoring, analysis, testing, and simulation in CSE. As discussed above, machine learning-supported and data-driven approaches can be used to (continuously) learn and tune the models. 

\subsubsection{Implications for practitioners and researchers}
\noindent
Practitioners can benefit from the presented results to understand how software performance engineering and software architectural can support practices in CSE and DevOps. Moreover, they can benefit from the classification of methodologies applicable to architectural support for performance engineering in the context of continuous software development.  

Researchers can benefit from our results to understand research trends and research gaps, and to better focus on their future work. In particular, the software performance community can leverage this work to understand whether their approaches are applicable in a CSE/DevOps context. Moreover, this study helps them define the requirements that drive the development of their tools to increase the chances of industrial adoption. In contrast, researchers in the field of CSE/DevOps can identify which methodologies and techniques can help improve software performance at some stage in the development cycle.

\smallskip
\noindent\fbox{%
    \parbox{\linewidth}{%
    \noindent
    \emph{Main findings:}
    \begin{itemize}[leftmargin=*]
        \item There is a growing interest for a culture of quality, where quality properties are continuously exposed to analysis and verification during the software development life-cycle;
        \item The complexity of the current performance engineering approaches should  improve their suitability for the DevOps principles of rapid changes, development automation, and fast feedback loop;
        \item There is a need for an established methodology for the systematically employment of performance engineering and machine learning/data-driven engineering in continuous development;
        \item Continuous monitoring and frequent re-assessment and re-architecting are necessary to reduce uncertainty in CSE and DevOps;
        \item Using higher abstraction level (i.e. by means of a model-based framework or (domain-specific) languages and automation mechanisms) can help in reducing the effort of integrating heterogeneous methods and components in systems. 
    \end{itemize}
    
    }%
}

%% file: threats.tex
\section{Threats to Validity}\label{sec:threats}
\noindent
Systematic Literature Review results might be affected by some threats mainly related to the correctness and completeness of the survey. In this section, we determined these threats according to the guidelines proposed by Wohlin et al.~\cite{Wohlin2012}: construct, internal, external, and conclusion validity threats. Moreover, we identified the actions required to mitigate them. 

\subsection{Construct validity}
\noindent
Construct validity is related to the generalization of the result to the concept or theory behind the execution of the study execution~\cite{Wohlin2012}. We identified the threats related to the potentially subjective analysis of the selected studies.
As recommended by the guidelines of Kitchenham ~\cite{Kitchenham2007}, data extraction was performed independently by two or more researchers and, in case of discrepancies, a third author was involved in the discussion to resolve any disagreement. The quality of each selected paper was checked according to the protocol proposed by Dyb{\aa} and Dings{\o}yr~\cite{Dyba2008}. 

\subsection{Internal validity}
\noindent
Internal validity threats are related to possible incorrect conclusions about the causal relationships between the  treatment and outcome ~\cite{Wohlin2012}. In the case of secondary studies, internal validity represents how well the findings represent those reported in literature.

To address these threats, we rigorously defined the study protocol, including the data-extraction form. The data extraction form was first validated by all authors by extracting information from 10 randomly selected papers. Considering the data analysis process, threats are minimal, as we only adopted descriptive statistical techniques when dealing with quantitative data.

When considering qualitative data, keywords were defined using a semi-automated approach to transform them into quantitative data. As regards to keyword definition, we first applied natural language techniques to reduce the subjectivity of the terms selected and then manually refined the keywords collaboratively.

Finally, 10 studies were randomly selected by all the researchers to verify whether the results were consistent, independent of the researcher performing the extraction. Disagreements were discussed and resolved collaboratively when needed.

\subsection{External validity}
\noindent
External validity threats are related to the ability to generalize the result~\cite{Wohlin2012}. In secondary studies, the external validity depends on the representativeness of the selected studies. If the selected studies are not externally valid, the synthesis of their content are not be valid. In our study, we were not able to evaluate the external validity of all the included studies.

To address this threat, we applied our search string to multiple bibliographic sources, including the Springer Link, Scopus, ACM Digital Library, and IEEEXplore Digital Library.
The usage of different bibliographic sources enabled us to guarantee to obtain the vast majority of papers.
Moreover, we also complemented our search by performing a snowballing activity.  
The inclusion of papers written only in English may have biased our results. Studies in other languages may be relevant. However, we have adopted English only as it is the language most widely for scientific papers, and we can consider the bias related to this threat as minimal. We only included peer-reviewed papers, without considering grey literature (e.g., technical reports, master theses, and web forums, etc.). Because we aimed to identify only high-quality scientific studies, we believed that this threat was minimal.

\subsection{Conclusion validity}
\noindent
Conclusion validity is related to the reliability of the conclusions drawn from the results~\cite{Wohlin2012}. 

One of these is related to the potential non-inclusion of some studies. To mitigate this threat, we carefully applied the search strategy and performed the search in eight digital libraries in conjunction with the snowballing process considering all the references presented in the retrieved papers and evaluating all the papers that reference the retrieved ones, which resulted in one additional relevant paper. We applied a broad search string, leading to a large set of articles and enabled us to include more possible results. We defined the inclusion and exclusion criteria and first applied them to the title and abstract. However, we did not rely exclusively on titles and abstracts, but before accepting a paper based on the title and abstract, we browsed the full text and applied our inclusion and exclusion criteria again.

Another possible conclusion validity threat is related to the incorrect interpretation of the results. To mitigate this threat, all authors carefully reviewed the results. However, other researchers may provide diﬀerent interpretations.

%% file: conclusion.tex
\section{Conclusion}\label{sec:conclusion}
 \noindent
 This paper presented a mapping study on the \topic. Of \initialPapers relevant studies, we selected \SelectedFullPapers primary studies, which were analyzed to answer our research questions. Thus, we have given a deeper look at the research context, and therefore we provided ideas to researchers and developers to address the challenges related to this topic, including the fact that knowledge gaps and future topics of research have not yet been thoroughly investigated in this context. In particular, we analyzed the publication trends, the research areas and target systems, target problems and contributions, and specific characteristics of the selected primary studies through a classification framework.

This study shows that SP and SA are aspects well considered in CSE, where the most affected dimensions are continuous monitoring and continuous improvement. The results of this study also show that SPE approaches and methodologies are suﬃciently mature (owing to the support of specific frameworks and tools) to be applied in continuous practices, with a prevalence in the use of data monitored at runtime. In general, SA is considered to oﬀer specific support; in many cases, SA models are used as input for the analysis and prediction of performance as well as architectural parameters and configurations. More support has been provided to distributed systems, component-based systems, SOA, micro services, and software intensive systems in general. Other contexts, such as data-intensive or embedded systems, have fewer applications. The most interesting gaps are identified in cloud systems and systems where uncertainty needs to be investigated.

%% file: primary_studies.tex
{\small
  \begin{enumerate}[labelindent=-5pt,label={[SP}{\arabic*]}]

\item\label{bib:Valetto2003} G. Valetto and G. Kaiser, Using process technology to control and coordinate software adaptation. International Conference on Software Engineering, 2003. \\ \url{https://doi.org/10.1109/ICSE.2003.1201206}
\item\label{bib:Garlan2004} Garlan D, Cheng SW, Huang AC, Schmerl B, Steenkiste P., Rainbow: Architecture-based self-adaptation with reusable infrastructure. Computer Journal, 2004. \\ \url{https://doi.org/10.1109/MC.2004.175}
\item\label{bib:DelRosso2006} Del Rosso, C., Continuous evolution through software architecture evaluation: A case study. Journal of Software Maintenance and Evolution, 2006. \\ \url{https://doi.org/10.1002/smr.337}
\item\label{bib:Tsai2006} T. Tsai and K. Vaidyanathan and K. Gross, Low-Overhead Run-Time Memory Leak Detection and Recovery. Pacific Rim International Symposium on Dependable Computing, 2006. \\ \url{https://doi.org/10.1109/PRDC.2006.42}
\item\label{bib:Liu2007} Y. Liu and I. Gorton and V. K. Le, A Configurable Event Correlation Architecture for Adaptive J2EE Applications. Australian Conference on Software Engineering, 2007. \\ \url{https://doi.org/10.1109/ASWEC.2007.5}
\item\label{bib:DelRosso2008} Del Rosso, C., Software performance tuning of software product family architectures: Two case studies in the real-time embedded systems domain. Journal of Systems and Software, 2008. \\ \url{https://doi.org/10.1016/j.jss.2007.07.006}
\item\label{bib:Epifani2009} I. Epifani and C. Ghezzi and R. Mirandola and G. Tamburrelli, Model evolution by run-time parameter adaptation. International Conference on Software Engineering, 2009. \\ \url{https://doi.org/10.1109/ICSE.2009.5070513}
\item\label{bib:Pooley2010} R. J. Pooley and A. A. L. Abdullatif, CPASA: Continuous Performance Assessment of Software Architecture. International Conference and Workshop on Engineering of Computer-Based Systems, 2010. \\ \url{https://doi.org/10.1109/ECBS.2010.16}
\item\label{bib:Brosig2011} Brosig F, Huber N, Kounev S., Automated extraction of architecture-level performance models of distributed component-based systems. International Conference on Automated Software Engineering, 2011. \\ \url{https://doi.org/10.1109/ASE.2011.6100052}
\item\label{bib:Ehlers2011a} Ehlers J, van Hoorn A, Waller J, Hasselbring W., Self-adaptive software system monitoring for performance anomaly localization. International Conference on Autonomic Computing, 2011. \\ \url{https://doi.org/10.1145/1998582.1998628}
\item\label{bib:Ehlers2011b} Ehlers, Jens and Hasselbring, Wilhelm, A Self-adaptive Monitoring Framework for Component-Based Software Systems. European Conference on Software Architecture, 2011. \\ \url{https://doi.org/10.1007/978-3-642-23798-0_30}
\item\label{bib:Calinescu2011} R. Calinescu, L. Grunske, M. Kwiatkowska, R. Mirandola and G. Tamburrelli, Dynamic qos management and optimization in service-based systems. IEEE Transactions on Software Engineering, 2011. \\ \url{https://doi.org/10.1109/TSE.2010.92}
\item\label{bib:RodrigoCalheiros2011} Rodrigo N. Calheiros, Rajiv Ranjan, Rajkumar Buyya, Virtual Machine Provisioning Based on Analytical Performance and QoS in Cloud Computing Environments. International Conference on Parallel Processing, 2011. \\ \url{https://doi.org/10.1109/ICPP.2011.17}
\item\label{bib:VonMassow2011} Von Massow, R. and Van Hoorn, A. and Hasselbring, W., Performance simulation of runtime reconfigurable component-based software architectures. European Conference on Software Architecture, 2011. \\ \url{https://doi.org/10.1007/978-3-642-23798-0_5}
\item\label{bib:Stochel2012} M. G. Stochel and M. R. Wawrowski and J. J. Waskiel, Adaptive Agile Performance Modeling and Testing. IEEE Annual Computer Software and Applications Conference Workshops, 2012. \\ \url{https://doi.org/10.1109/COMPSACW.2012.85}
\item\label{bib:vonLaszewski2012} von Laszewski, Gregor and Lee, Hyungro and Diaz, Javier and Wang, Fugang and Tanaka, Koji and Karavinkoppa, Shubhada and Fox, Geoffrey C. and Furlani, Tom, Design of an Accounting and Metric-Basedcloud-Shifting and Cloud-Seeding Framework for Federatedclouds and Bare-Metal Environments. workshop on Cloud services, federation, and the 8th open cirrus summit, 2012. \\ \url{https://doi.org/10.1145/2378975.2378982}
\item\label{bib:ReinerJung2013} Reiner Jung, Robert Heinrich, Eric Schmieders, Model-driven instrumentation with Kieker and Palladio to forecast dynamic applications. Proc. Kieker/Palladio Days, 2013. \\ \url{http://ceur-ws.org/Vol-1083/paper11.pdf}
\item\label{bib:Wert2013} Wert, Alexander, Performance Problem Diagnostics by Systematic Experimentation. CompArch: Component-Based Software Engineering and Software Architecture, 2013. \\ \url{https://doi.org/10.1145/2465498.2465499}
\item\label{bib:Chiprianov2014} Chiprianov V, Falkner K, Szabo C, Puddy G., Architectural support for model-driven performance prediction of distributed real-time embedded systems of systems. European Conference on Software Architecture, 2014. \\ \url{https://doi.org/10.1007/978-3-319-09970-5_30}
\item\label{bib:Junior2014} M. S. S. Junior and N. S. Rosa and F. A. A. Lins, Execution Support to Long Running Workflows. International Conference on Computer and Information Technology, 2014. \\ \url{https://doi.org/10.1109/CIT.2014.94}
\item\label{bib:Birngruber2015} Birngruber, Erich and Forai, Petar and Zauner, Aaron, Total Recall: Holistic Metrics for Broad Systems Performance and User Experience Visibility in a Data-Intensive Computing Environment. International Workshop on HPC User Support Tools, 2015. \\ \url{https://doi.org/10.1145/2834996.2835001}
\item\label{bib:Incerto2015} Incerto, E. and Tribastone, M. and Trubiani, C., A proactive approach for runtime self-adaptation based on queueing network fluid analysis. 1st International Workshop on Quality-Aware DevOps, QUDOS 2015, 2015. \\ \url{https://doi.org/10.1145/2804371.2804375}
\item\label{bib:Wert2015} Wert, Alexander and Schulz, Henning and Heger, Christoph, AIM: Adaptable Instrumentation and Monitoring for Automated Software Performance Analysis. International Workshop on Automation of Software Test, 2015. \\ \url{https://doi.org/10.1109/AST.2015.15}
\item\label{bib:Brebner2016} Brebner, P., Automatic performance modelling from application performance management (APM) data: An experience report. International Conference on Performance Engineering, 2016. \\ \url{https://doi.org/10.1145/2851553.2851560}
\item\label{bib:Gerostathopoulos2016} Gerostathopoulos, I. and Bures, T. and Schmid, S. and Horky, V. and Prehofer, C. and Tuma, P., Towards systematic live experimentation in software-intensive systems of systems. International Colloquium on Software-intensive Systems-of-Systems, 2016. \\ \url{https://doi.org/10.1145/3175731.3176175}
\item\label{bib:Heinrich2016} Heinrich R., Architectural Run-time Models for Performance and Privacy Analysis in Dynamic Cloud Applications. Performance Evaluation, 2016. \\ \url{https://doi.org/10.1145/2897356.2897359}
\item\label{bib:Incerto2016} Incerto, E. and Tribastone, M. and Trubiani, C., Symbolic performance adaptation. International Symposium on Software Engineering for Adaptive and Self-Managing Systems, 2016. \\ \url{https://doi.org/10.1145/2897053.2897060}
\item\label{bib:Keck2016} Keck, P. and Hoorn, A.V. and Okanovic, D. and Pitakrat, T. and Dullmann, T.F., Antipattern-Based Problem Injection for Assessing Performance and Reliability Evaluation Techniques. International Conference on Software Reliability Engineering Workshops, 2016. \\ \url{https://doi.org/10.1109/ISSREW.2016.36}
\item\label{bib:Spinner2016} Spinner S, Walter J, Kounev S., A Reference Architecture for Online Performance Model Extraction in Virtualized Environments. Workshop on Challenges in Performance Methods for Software Development, 2016. \\ \url{https://doi.org/10.1145/2859889.2859893}
\item\label{bib:Willnecker2016} Willnecker, F. and Krcmar, H., Optimization of deployment topologies for distributed enterprise applications. International ACM SIGSOFT Conference on Quality of Software Architectures, 2016. \\ \url{https://doi.org/10.1109/QoSA.2016.11}
\item\label{bib:Brunnert2017} Brunnert A, Krcmar H., Continuous performance evaluation and capacity planning using resource profiles for enterprise applications. Journal of Systems and Software, 2017. \\ \url{https://doi.org/10.1016/j.jss.2015.08.030}
\item\label{bib:Incerto2017} Incerto, E. and Tribastone, M. and Trubiani, C., Software performance self-adaptation through efficient model predictive control. International Conference on Automated Software Engineering, 2017. \\ \url{https://doi.org/10.1109/ASE.2017.8115660}
\item\label{bib:Kunz2017} Kunz, J. and Heger, C. and Heinrich, R., A generic platform for transforming monitoring data into performance models. Companion International Conference on Performance Engineering, 2017. \\ \url{https://doi.org/10.1145/3053600.3053635}
\item\label{bib:Li2017} Li, Chenand Altamimi, Taghreedand Zargari, Mana Hassanzadehand Casale, Giulianoand Petriu, Dorina, Tulsa: A Tool for Transforming UML to Layered Queueing Networks for Performance Analysis of Data Intensive Applications. Quantitative Evaluation of Systems, 2017. \\ \url{https://doi.org/10.1007/978-3-319-66335-7_18}
\item\label{bib:Perez-Palacin2017} Perez-Palacin D, Ridene Y, Merseguer J., Quality assessment in DevOps: automated analysis of a tax fraud detection system. International Workshop on Quality-aware DevOps, 2017. \\ \url{https://doi.org/10.1145/3053600.3053632}
\item\label{bib:Trubiani2017} Trubiani, C. and Mirandola, R., Continuous rearchitecting of QoS models: Collaborative analysis for uncertainty reduction. European Conference on Software Architecture, 2017. \\ \url{https://doi.org/10.1007/978-3-319-65831-5_3}
\item\label{bib:Walter2017} Walter, J. and Stier, C. and Koziolek, H. and Kounev, S., An expandable extraction framework for architectural performance models. Companion International Conference on Performance Engineering, 2017. \\ \url{https://doi.org/10.1145/3053600.3053634}
\item\label{bib:Bao2017} Y. Bao and M. Chen and Q. Zhu and T. Wei and F. Mallet and T. Zhou, Quantitative Performance Evaluation of Uncertainty-Aware Hybrid AADL Designs Using Statistical Model Checking. IEEE Transactions on Computer-Aided Design of Integrated Circuits and Systems, 2017. \\ \url{https://doi.org/10.1109/TCAD.2017.2681076}
\item\label{bib:Aleti2018} Aleti, A. and Trubiani, C. and van Hoorn, A. and Jamshidi, P., An efficient method for uncertainty propagation in robust software performance estimation. Journal of Systems and Software, 2018. \\ \url{https://doi.org/10.1016/j.jss.2018.01.010}
\item\label{bib:Bernardi2018} Bernardi, Simona and Dominguez, Juan L. and Gomez, Abel and Joubert, Christophe and Merseguer, Jose and Perez-Palacin, Diego and Requeno, Jose I. and Romeu, Alberto, A systematic approach for performance assessment using process mining. Empirical Software Engineering, 2018. \\ \url{https://doi.org/10.1007/s10664-018-9606-9}
\item\label{bib:Cholomskis2018} Cholomskis, Aurimasand Pozdniakova, Olesiaand Mažeika, Dalius, Cloud Software Performance Metrics Collection and Aggregation for Auto-Scaling Module. International Conference on Information and Software Technologies, 2018. \\ \url{https://doi.org/10.1007/978-3-319-99972-2_10}
\item\label{bib:Bardsley2018} D. Bardsley and L. Ryan and J. Howard, Serverless Performance and Optimization Strategies. International Conference on Smart Cloud, 2018. \\ \url{https://doi.org/10.1109/SmartCloud.2018.00012}
\item\label{bib:Falkner2018} Falkner, Katrina and Szabo, Claudia and Chiprianov, Vanea and Puddy, Gavin and Rieckmann, Marianne and Fraser, Dan and Aston, Cathlyn, Model-driven performance prediction of systems of systems. Software Systems Modeling, 2018. \\ \url{https://doi.org/10.1007/s10270-016-0547-8}
\item\label{bib:Pitakrat2018} Pitakrat, T. and Okanović, D. and van Hoorn, A. and Grunske, L., Hora: Architecture-aware online failure prediction. Journal of Systems and Software, 2018. \\ \url{https://doi.org/10.1016/j.jss.2017.02.041}
\item\label{bib:Trubiani2018} Trubiani, C. and Bran, A. and van Hoorn, A. and Avritzer, A. and Knoche, H., Exploiting load testing and profiling for Performance Antipattern Detection. Information and Software Technology, 2018. \\ \url{https://doi.org/10.1016/j.infsof.2017.11.016}
\item\label{bib:Vögele2018} Vögele, Christianand van Hoorn, Andréand Schulz, Eikeand Hasselbring, Wilhelmand Krcmar, Helmut, WESSBAS: extraction of probabilistic workload specifications for load testing and performance prediction---a model-driven approach for session-based application systems. Software Systems Modeling, 2018. \\ \url{https://doi.org/10.1007/s10270-016-0566-5}
\item\label{bib:Willnecker2018} Willnecker, F. and Krcmar, H., Multi-objective optimization of deployment topologies for distributed applications. ACM Transactions on Internet Technology, 2018. \\ \url{https://doi.org/10.1145/3106158}
\item\label{bib:Arcelli2019} Cortellessa, V., Di Pompeo, D., Eramo, R., Tucci, M., A model-driven approach for continuous performance engineering in microservice-based systems.	Journal of Systems and Software, 2022. \\ \url{https://doi.org/10.1016/j.jss.2021.111084}	
\item\label{bib:Bezemer2019} Bezemer, C.-P. and Eismann, S. and Ferme, V. and Grohmann, J. and Heinrich, R. and Jamshidi, P. and Shang, W. and Van Hoorn, A. and Villavicencio, M. and Walter, J. and Willnecker, F., How is performance addressed in DevOps? A survey on industrial practices. International Conference on Performance Engineering, 2019. \\ \url{https://doi.org/10.1145/3297663.3309672}
\item\label{bib:Castellanos2019} Castellanos, C. and Varela, C.A. and Correal, D., Measuring performance quality scenarios in big data analytics applications: A DevOps and domain-specific model approach. ACM International Conference Proceeding Series, 2019. \\ \url{https://doi.org/10.1145/3344948.3344986}
\item\label{bib:Grohmann2019} Grohmann, J. and Eismann, S. and Elflein, S. and Kistowski, J.V. and Kounev, S. and Mazkatli, M., Detecting parametric dependencies for performance models using feature selection techniques. Annual International Symposium on Modeling, Analysis, and Simulation of Computer and Telecommunications Systems, MASCOTS, 2019. \\ \url{https://doi.org/10.1109/MASCOTS.2019.00042}
\item\label{bib:Muller2019} Muller, H. and Bosse, S. and Turowski, K., On the utility of machine learning for service capacity management of enterprise applications. International Conference on Signal Image Technology and Internet Based Systems, 2019. \\ \url{https://doi.org/10.1109/SITIS.2019.00053}
\item\label{bib:Perez-Palacin2019} Perez-Palacin, D. and Merseguer, J. and Requeno, J.I. and Guerriero, M. and Di Nitto, E. and Tamburri, D.A., A UML Profile for the Design, Quality Assessment and Deployment of Data-intensive Applications. Software and Systems Modeling, 2019. \\ \url{https://doi.org/10.1007/s10270-019-00730-3}
\item\label{bib:Chatley2019} R. Chatley and T. Field and D. Wei, Continuous Performance Testing in Virtual Time. International Conference on Software Architecture Workshops, 2019. \\ \url{https://doi.org/10.1109/ICSA-C.2019.00027}
\item\label{bib:Spinnner2019} Spinnner, S. and Grohmann, J. and Eismann, S. and Kounev, S., Online model learning for self-aware computing infrastructures. Journal of Systems and Software, 2019. \\ \url{https://doi.org/10.1016/j.jss.2018.09.089}
\item\label{bib:Trubiani2019} Trubiani, C. and Jamshidi, P. and Cito, J. and Shang, W. and Jiang, Z.M. and Borg, M., Performance issues? Hey DevOps, mind the uncertainty. IEEE Software, 2019. \\ \url{https://doi.org/10.1109/MS.2018.2875989}
\item\label{bib:Willnecker2019} Willnecker, F. and Krcmar, H., Model-based prediction of automatic memory management and garbage collection behavior. Simulation Modelling Practice and Theory, 2019. \\ \url{https://doi.org/10.1016/j.simpat.2018.09.014}
\item\label{bib:Yasaweerasinghelage2019} Yasaweerasinghelage, R. and Staples, M. and Paik, H.-Y. and Weber, I., Optimising architectures for performance, cost, and security. European Conference on Software Architecture, 2019. \\ \url{https://doi.org/10.1007/978-3-030-29983-5_11}
\item\label{bib:DeSanctis2020} De Sanctis, M. and Bucchiarone, A. and Trubiani, C., A DevOps Perspective for QoS-Aware Adaptive Applications. International Workshop on Software Engineering Aspects of Continuous Development and New Paradigms of Software Production and Deployment, 2020. \\ \url{https://doi.org/10.1007/978-3-030-39306-9_7}
\item\label{bib:Giaimo2020} Giaimo, Federico and Berger, Christian, Continuous Experimentation for Automotive Software on the Example of a Heavy Commercial Vehicle in Daily Operation. European Conference on Software Architecture, 2020. \\ \url{https://doi.org/10.1007/978-3-030-58923-3_5}
\item\label{bib:Heinrich2020} Heinrich, R., Architectural runtime models for integrating runtime observations and component-based models. Journal of Systems and Software, 2020. \\ \url{https://doi.org/10.1016/j.jss.2020.110722}
\item\label{bib:Mazkatli2020} Mazkatli, M. and Monschein, D. and Grohmann, J. and Koziolek, A., Incremental calibration of architectural performance models with parametric dependencies. International Conference on Software Architecture, 2020. \\ \url{https://doi.org/10.1109/ICSA47634.2020.00011}
\item\label{bib:Voneva2020} Voneva, Sonya and Mazkatli, Manar and Grohmann, Johannes and Koziolek, Anne, Optimizing Parametric Dependencies for Incremental Performance Model Extraction. European Conference on Software Architecture, 2020. \\ \url{https://doi.org/10.1007/978-3-030-59155-7_17}
\item\label{bib:Camilli2022} Camilli, M., Janes, A., Russo, B., Automated test-based learning and verification of performance models for microservices systems, Journal of Systems and Software, 2022. \url{https://doi.org/10.1016/j.jss.2022.111225}
\item\label{bib:Kugele2021}Kugele, S., Obergfell, P., Sax, E., Model-based resource analysis and synthesis of service-oriented automotive software architectures, Software and Systems Modeling, 2021. \\ \url{https://doi.org/10.1007/s10270-021-00896-9}
\item\label{bib:Grohmann2021}Grohmann, J., Eismann, S., Bauer, A., Spinner, S., Blum, J., Herbst, N., Kounev, S., SARDE: A Framework for Continuous and Self-Adaptive Resource Demand Estimation, ACM Transactions on Autonomous and Adaptive Systems, 2021.\\ \url{https://doi.org/10.1145/3463369}
\end{enumerate}
}

%% file: table_data_extraction.tex

\begin{table*}
\centering
\rowcolors{2}{gray!25}{white}
\resizebox{1\textwidth}{!}{
\begin{tabular}{c|cccccc|ccccccc|ccccc|ccccccc|cccccccccc|cccccc|cccccc} 
\hline
\multirow{2}{*}{} & \multicolumn{6}{c|}{Research area} & \multicolumn{7}{c|}{Target systems}& \multicolumn{5}{c|}{Research problems} & \multicolumn{7}{c|}{Research contributions} & \multicolumn{10}{c|}{Methodologies/techniques} & \multicolumn{6}{c|}{Measures/indices} & \multicolumn{6}{c|}{Data} \\
\hline
\end{tabular}
}
\caption{Data extraction results}\label{tab:data-extraction-results}
\end{table*}

%% file: table_keywords.tex
\begin{scriptsize}

\begin{table*}[!h]
%
\caption{Research areas - keywords} \label{tab:reserach-keywords}
\end{table*}

\begin{table*}[]
%
\caption{Target systems - keywords} \label{tab:domanis-keyword}
\end{table*}

\begin{table*}
%
\caption{Target problems - keywords} \label{tab:target-keywords}
\end{table*}

\begin{table*}
%
\caption{Contributions - keywords} \label{tab:contribution-keywords}
\end{table*}

\begin{table*}
%
\caption{Methodologies - keywords} \label{tab:methodology-keyworkds}
\end{table*}

\begin{table*}
%
\caption{Measures/Indices - keywords} \label{tab:measures-keywords}
\end{table*}

\begin{table*}
%
  \\ 
\bottomrule
\end{tabular}
\caption{Data - keywords} \label{tab:data-keywords}
\end{table*}

\end{scriptsize}